\documentclass[a4paper,12pt]{article}
\usepackage{a4}
\usepackage{amssymb}
\usepackage{amsmath}
\usepackage{epsfig}

\begin{document}

%macros
\newcommand{\todo}[1]{{\em \small {#1}}\marginpar{$\Longleftarrow$}}
\newcommand{\labell}[1]{\label{#1}\qquad_{#1}} %{\label{#1}} %
\newcommand{\bbibitem}[1]{\bibitem{#1}\marginpar{#1}}
\newcommand{\llabel}[1]{\label{#1}\marginpar{#1}}
\newcommand{\restrict}[1]{\lfloor_{#1}}
\renewcommand{\binom}[2]{\left(#1 \atop #2\right)}

\newtheorem{theorem}{Theorem}
\newtheorem{lemma}{Lemma}
\newcommand{\tr}{\mathrm{tr}}
\newcommand{\xxx}{$X\!X\!X\,$}
\newcommand{\be}{\begin{equation}}
\newcommand{\ee}{\end{equation}}

\hyphenation{Min-kow-ski}
\hyphenation{four-di-men-sio-nal}

\def\eps{\epsilon}

\def\NPB{{\it Nucl. Phys. }{\bf B}}
\def\PL{{\it Phys. Lett. }}
\def\PRL{{\it Phys. Rev. Lett. }}
\def\PRD{{\it Phys. Rev. }{\bf D}}
\def\CQG{{\it Class. Quantum Grav. }}
\def\JMP{{\it J. Math. Phys. }}
\def\SJNP{{\it Sov. J. Nucl. Phys. }}
\def\SPJ{{\it Sov. Phys. J. }}
\def\JETPL{{\it JETP Lett. }}
\def\TMP{{\it Theor. Math. Phys. }}
\def\IJMPA{{\it Int. J. Mod. Phys. }{\bf A}}
\def\MPL{{\it Mod. Phys. Lett. }}
\def\CMP{{\it Commun. Math. Phys. }}
\def\AP{{\it Ann. Phys. }}
\def\PR{{\it Phys. Rep. }}

\renewcommand{\thepage}{\arabic{page}}
\setcounter{page}{1}

%\title
\rightline{hep-th/0411110}
\rightline{ITFA-2004-56}
\vskip 0.75 cm
\renewcommand{\thefootnote}{\fnsymbol{footnote}}
\begin{center}
\Large \bf Classical Spin Chains and Exact Three-dimensional Superpotentials
\end{center}
\vskip 0.75 cm

\centerline{{\bf Rutger Boels\footnote{rhboels@science.uva.nl} and 
Jan de Boer,\footnote{jdeboer@science.uva.nl}
}}
\vskip .5cm
\centerline{\it Instituut voor Theoretische Fysica,}
\centerline{\it Valckenierstraat 65, 1018XE Amsterdam, The Netherlands}
\vskip .5cm

\setcounter{footnote}{0}
\renewcommand{\thefootnote}{\arabic{footnote}}

\begin{abstract}
\noindent We study exact effective superpotentials of four-dimensional \mbox{${\cal N}=2$} supersymmetric gauge theories with gauge group $U(N)$ and various amounts of fundamental matter on \mbox{$\mathbb{R}^3\times S^1$}, broken to ${\cal N}=1$ by turning on a classical superpotential for the adjoint scalar. On general grounds these superpotentials can easily be constructed once we identify a suitable set of coordinates on the moduli space of the gauge theory. These coordinates have been conjectured to be the phase space variables of the classical integrable system which underlies the \mbox{${\cal N}=2$} gauge theory.  The sought low energy effective superpotential can then be constructed from the conserved quantities in the integrable system. For the gauge theory under study these integrable systems are degenerations of the classical, inhomogeneous, periodic $SL(2, \mathbb{C})$ spin chain. Ambiguities in the degeneration provide multiple coordinate patches on the gauge theory moduli space. By studying the vacua of these superpotentials in several examples we find that the spin chain provides coordinate patches that parametrize holomorphically the part of the gauge theory moduli space which is connected to the electric (as opposed to magnetic or baryonic) Higgs and Coulomb branch vacua. The baryonic branch root is on the edge of some coordinate patches.  As a product of our analysis all maximally confining (non-baryonic) Seiberg-Witten curve factorizations for $N_f \leq N_c$ are obtained, explicit up to one constraint for equal mass flavors and up to two constraints for unequal mass flavors. Gauge theory addition and multiplication maps are shown to have a natural counterpart in this construction. Furthermore it is shown how to integrate in the meson fields in this formulation in order to obtain three and four dimensional Affleck-Dine-Seiberg-like superpotentials.
\end{abstract}
\newpage

\section{Introduction}

The determination of the precise structure of gauge theories in
the infra-red is an important and elusive problem. Recently, much
progress has been made in calculating the exact superpotentials in
the low energy effective action in a large class of minimally
supersymmetric ($\mathcal{N} = 1$) four-dimensional gauge
theories. These theories are obtained by breaking $\mathcal{N} =
2$ to $\mathcal{N} =1$ by turning on a classical superpotential.
One way to study these theories involves the relation between
$\mathcal{N}=2$ theories and integrable systems: the
Seiberg-Witten curve of $\mathcal{N}=2$ theories can be identified
with the spectral curve of an underlying integrable system. This identification relates the gauge theory moduli and the conserved charges (action variables) of the integrable system. If the gauge theory is compactified on  \mbox{$\mathbb{R}^3\times S^1$},
the relationship between gauge theory and integrable system
becomes tighter, since the extra moduli from the compactification
can be identified with the angle variables of the integrable
system.

Seiberg and Witten \cite{seibergwittencompac} have shown that the
moduli spaces of four dimensional $\mathcal{N} = 2$ gauge theories
on  \mbox{$\mathbb{R}^3\times S^1$} have a distinguished complex
structure which is independent of the radius of the $S^1$. The
superpotential, which is a holomorphic quantity, will therefore
also be independent of the radius of the $S^1$, and the same is true for the vacuum structure of the theory. They do however change in the three-dimensional limit $R \rightarrow 0$, since in that case we should work with the 3d gauge coupling instead of the 4d scale $\Lambda$, and the relation between the two explicitly involves R. In this paper we restrict attention to finite $R$, as the generalization to zero radius is rather straightforward.

Though the vacuum structure of the gauge theory on $\mathbb R^4$
is the same as that on $\mathbb{R}^3 \times S^1$, the field content
and in particular the superpotential of each are quite distinct.
There is a significant advantage to working on $\mathbb{R}^3 \times
S^1$ compared to $\mathbb R^4$. The four dimensional gauge field
gives, after compactification on $S^1$, rise to a pair of scalar
fields in three dimensions. We can give these scalar fields a
vacuum expectation value which allows us to study the theory
effectively at weak coupling. In particular, all non-perturbative
effects will be due to conventional three-dimensional instantons.
This in contrast to the four-dimensional situation where the
non-perturbative physics is due to more complicated gauge field
configurations such as fractional instantons.

Based on these considerations and on previous work \cite{dorey1, dorey2, dorey3}, a precise conjecture was made in \cite{firstpaper} about the sought exact superpotential on $ \mathbb{R}^3 \times S^1$: it can be obtained by replacing the gauge invariant operators appearing in the classical superpotential
by corresponding conserved quantities of the underlying integrable system. To achieve this, one
can simply replace the adjoint superfield appearing in the superpotential by the Lax matrix of
the integrable system. This is a very suggestive operation reminiscent of a master field, though
a precise interpretation along these lines has not yet been found.
In \cite{firstpaper} this conjecture was verified by calculating the extrema of the
resulting superpotential in various examples, and a precise agreement with known results in
four dimensions was found, and in particular the Seiberg-Witten curve always factorized in the
appropriate way. A general proof was given in \cite{secondpaper, hollowood}.
Several generalizations have been made since, to $\cal N$ $=1^*$ \cite{hollowood},
to other classical gauge groups \cite{mohsen} and even to theories with
gauge group $G_2$ \cite{thirdpaper}. In \cite{hollowoodxxx} $SU(2)$ with four flavours was studied.

In this paper we discuss the generalization to $U(N)$ gauge theories with $N_f$ hypermultiplets
in the fundamental representation of the gauge group. According to \cite{xxxrussians},
the integrable system for $N_f < 2 N_c$ should be a degeneration of the
classical version of the inhomogeneous \xxx spin chain. It was shown
in that paper that the form of the spectral curve (for $N_f = 2 N_c$)
matched that of the Seiberg-Witten curve as in \cite{hanany}. Quite generally, any holomorphic integrable system with the property that (i) the conserved charges (the action variables) precisely parametrize the moduli space of the four-dimensional gauge theory, (ii) the spectral curve always agrees with the Seiberg-Witten curve, and (iii) the angle variables parametrize the Jacobian of the Seiberg-Witten curve, is a candidate integrable system that one can use to describe the quantum superpotential of the gauge theory on ${\mathbb R}^3 \times S^1$. In this case the canonical variables (e.g. the coordinates and momenta) of the integrable system are complex coordinates on the moduli space of the gauge theory on ${\mathbb R}^3 \times S^1$, and the quantum superpotential is simply an appropriate linear combination of the conserved charges. Thus, the main purpose of the integrable system is to provide a suitable set of complex coordinates on the moduli space. Unfortunately, these coordinates are not generically good global coordinates. If they were, the equations that put the conserved charges equal to fixed numbers would define a torus (the Jacobian) embedded in the phase space of the integrable system. Except for the one-dimensional torus, which can be written as a cubic equation in $\mathbb C^2$, the equations that describe complex embeddings of complex tori are very complicated. Therefore, what we will find is that the integrable systems parametrize an open subset of the Jacobian but not the full moduli space of the gauge theory on $\mathbb{R}^3 \times S^1$. Correspondingly, we will see that some vacua of the gauge theory are not captured by the integrable system, or that they live at the boundary of phase space. Curiously enough, the 'electric' vacua of the theory are typically all recovered, while the dual `magnetic' vacua are not. Another phenomenon that we will encounter is that there are several different integrable systems all describing the gauge theory with $N_f<2 N_c$. These arise because there are several different ways in which we can degenerate the \xxx spin chain. The different integrable systems parametrize different open subsets of the Jacobian, and on the overlap the relation between them is given by a suitable canonical transformation between the phase space variables.

This paper is structured as follows: In section \ref{sec:gaugeexp} we discuss the gauge theory expectations for the vacuum structure of theories with matter in the fundamental representation. Section \ref{sec:spinchain} proceeds by introducing the classical spin chain, its spectral curve and its degenerations (down to eventually Toda). The following section, \ref{sec:proposal}, describes the natural proposal for the exact superpotential.  The addition and multiplication maps are discussed in section~\ref{sec:admultmaps}. Section \ref{sec:examples} contains the result of several example calculations and some words to clarify their meaning. Specifically, we show how particular degenerations can sometimes lead to extra vacua. In addition, here all massive (field theory) vacua for a generic superpotential are obtained. In section \ref{sec:integrmesons} it is shown how to integrate in meson fields into our proposed superpotential. We end with a discussion and present some conclusions and open problems. In several appendices we present some technical details, including the resolution of a point of minor confusion in the literature for high order superpotentials and an observation.  

\section{Gauge theory expectations}
\label{sec:gaugeexp}

In this section we will very briefly review the zoo of phases of
(the low energy effective theory corresponding to) $\cal N$ $= 2$
$U(N_c)$ 4 dimensional gauge theories with $N_f$ ($\leq 2 N_c$)
fundamental matter hypermultiplets, softly broken to $\mathcal{N}
= 1$ by turning on a superpotential (the literature on this
subject is vast, see e.g. \cite{argyresseiberg},
\cite{carlinokonishi} or the review in \cite{balastonaqvi} and
references therein). 

\subsection{Supersymmetric gauge theories on $\mathbb{R}^4$}
The perturbative field content of the $\mathcal{N} = 2$
supersymmetric gauge theory we will be studying consists of the
$U(N)$ gauge vector multiplet, which contains an ${\cal N}=1$
adjoint scalar superfield $\Phi$,\footnote{We will denote the
scalar components of the multiplets by the same symbol} and $N_f$
pairs of massive hypermultiplets $(\tilde{Q}, Q)$ in the (anti)
fundamental. The global classical symmetries are an $SU(2) \times
U(1)$ R-symmetry and an $SU(N_f) \times SU(N_f) \times U(1)_B$
flavour symmetry in the limit of vanishing masses. The
superpotential for theories with ${\cal N}=2$ supersymmetry and
fundamental matter is\footnote{With \cite{argyresseiberg} we
follow the conventions of \cite{wesbagger}.}
\begin{equation}
\mathcal{W}_{\mathcal{N} = 2} = \sqrt{2} \tilde{Q}_i^a \Phi_a^b Q_b^i + \sqrt{2} \tilde{Q}_i^a m_j^i Q_a^j,
\label{N2superpotential} \end{equation}
\begin{equation}
m_i^j = \textrm{Diag}(m_1, m_2 \ldots, m_{N_f}).
\end{equation}
It is well known that this theory is perturbatively asymptotically
free in the UV for $N_f < 2 N_c$ and finite for $N_f = 2 N_c$. We
will break $\mathcal{N} = 2$ softly to $\mathcal{N} = 1$ by
turning on a polynomial superpotential $W_{n+1}(\phi) =
\sum_{k=0}^{n+1} \frac{g_k}{k} \tr(\Phi^k)$ for the $\mathcal{N} =
1$ adjoint scalar $\Phi$. These theories have a rich vacuum structure, which can be studied by a variety of methods. Below we
briefly present the results of a field theoretical analysis. The
first step is calculating the minima of the classical gauge
theory, in order to identify the possible phases of the theory.
Since some of them will not receive quantum corrections the
results there are exact. The D-term equations of the theory read
\begin{eqnarray}
[\Phi,\Phi^\dagger] & = & 0 \nonumber \\
Q^i_a (Q^{\dagger})_i^b - \tilde{Q}^i_a (\tilde{Q}^{\dagger})_i^b & = & \nu \delta_a^b, \qquad \nu \in \mathbb{R}.
\label{Dterms}
\end{eqnarray}
with $\nu$ the Fayet-Illiopoulos parameter, which in almost all
cases must be zero in order for the classical theory to have a
supersymmetric vacuum. We will not consider non-zero $\nu$ in this
paper. The F-terms give
\begin{eqnarray}
W'(\Phi)_a^b + \sqrt{2} \tilde{Q}_i^a Q_b^i & = & 0\\
\Phi_a^b Q_b^i + m_j^i Q_a^j & = & 0\\
\tilde{Q}_i^b \Phi_b^a + \tilde{Q}_i^a m_j^i & = & 0
\label{Fterms}\end{eqnarray} As is well known, the first D-term
equation states that gauge transformations can be used to
diagonalize $\Phi = \textrm{Diag}(\phi_1, \phi_2, \phi_3, \ldots
\phi_{N_c})$. The F-terms state that the quark fields are either
zero or eigenvectors of $\Phi$ with $\phi_a = - m_i$ for some $a$
and $i$. If the quark masses are all distinct, we can label the
quark fields and the adjoint scalar eigenvalues for which this
condition is satisfied by $a=i=1,\ldots, r$. For every $r$ there
are $\binom{N_f}{r}$ ways of choosing masses for which this
condition holds. For unequal masses there are then $r$ distinct
eigenvectors of $\Phi$ which we can group into
diagonal\footnote{by diagonal we mean $a_{ij} =0$ if $i \neq j$
and if $i,j> \textrm{min}(N_c, N_f)$} $N_c \times N_f$ matrices for $Q$ and
$\tilde{Q}^t$,
\begin{eqnarray}
Q & =& \textrm{Diag}(Q_1, \ldots, Q_r, 0, \ldots)\\
\tilde{Q}^t & =& \textrm{Diag}(\tilde{Q}_1, \ldots, \tilde{Q}_r, 0, \ldots).
\end{eqnarray}
If some masses happen to be equal, than instead of a diagonal $Q$
matrix we get a block diagonal matrix $Q$ with sub-blocks $Q_i$, but we will first assume
unequal masses. The remaining D and F term equations now split
into equations for the different diagonal elements in $Q$. We get
\begin{equation}
|Q_i|^2 = |\tilde{Q}_i|^2.
\label{eq:modconstraint}
\end{equation}
Next, we order the masses and the eigenvalues of the adjoint
scalar such that the first $r$ of each are equal. The value of the
quark vevs and the remaining eigenvalues of the adjoint scalar can
be determined from the remaining F term equations,
\begin{eqnarray}
W'(-m_i) + \sqrt{2} \tilde{Q}_i Q^i  =  0, & & \quad i=1, \ldots, r \quad \textrm{(no sum)} \\ \label{eq:remaining}
W'(\phi_a) = 0, & & \quad a= r+1, \ldots, N_c.
\end{eqnarray}
If $W'(-m_i)$ is zero for some $i$, the quark fields $Q_i$ should
be zero and do not give rise to a Higgs branch. If $W'(-m_i)$ is
non-zero, the quark gets a vev which Higgses the corresponding
$U(1)$ gauge group. Clearly, the number of vacua depends on the
detailed form of the superpotential. For example, a quadratic
superpotential will force all the $\tilde{N}=N_c-r$ eigenvalues of
the adjoint scalar $\phi_a$ which are not equal to some $-m_i$ to
be equal to each other. The resulting $U(\tilde{N})$ theory is known to have
$\tilde{N}$ supersymmetric vacua.

When some masses are equal (as they will be in most of the cases
we consider) more elaborate vacua are possible. See the analysis
in, for instance, \cite{argyresseiberg}. In summary the following
happens: If $N_i$ eigenvalues of the adjoint become fixed at one
mass we retain a $U(N_i)$ gauge theory with a number $M_i$ of
massless flavors. If $W'(-m_i)$ is not zero, the combination of
equations (\ref{eq:modconstraint}) and (\ref{eq:remaining})
(suitably generalized for matrices $Q$ and $\tilde{Q}$) forces $Q$
and $\tilde{Q}$ to acquire $N_i$ non-zero eigenvalues which
completely Higgs the $U(N_i)$ theory. If $W'(-m_i)$ is zero,  at
least some $Q$ and $\tilde{Q}$ must be zero (and, in fact, they
may all be). The solutions are then labeled by an index $l \leq
\min(\frac{M_i}{2}, N_i)$ and generically the quark fields both
have $l$ degrees of freedom, constrained by the remaining D term
equation. The theory that described the remaining degrees of
freedom is then a $U(N_i-l)$ gauge theory with $M_i-2 l$ massless
flavors and $l(M_i -l)$ massless neutral Goldstone hypermultiplets
from the broken flavor symmetry.

Summarizing, the theory has several phases, which we distinguish
by the expectation values of the quark fields. If these are zero,
then we are on the Coulomb branch. If they are nonzero then we are
on the Higgs branch. It can happen that the expectation values of
the quark fields on this branch can be tuned to zero, in this case
we can continuously interpolate between Coulomb and Higgs branch.
The special point on which they connect is the root of the Higgs branch, and in order
for it to touch the Coulomb branch classically, $W'(-m_i)$ must be
zero. A question is whether or not this last point might be
modified in the quantum theory: while the metric on the Higgs
branch does not get quantum corrected, this does not mean that the
(position of the) boundary of moduli space can not get quantum
corrections. In the study of the integrable system we usually take
$W'(-m)$ to be zero, however we have not found a special status of
this extra condition on the superpotential: the Higgs branch continues
to be there in the analysis.

\subsubsection{Quantum mechanics}
As is well known, the moduli space of supersymmetric gauge theory
is governed by holomorphy and the non-renormalization theorems. The
metric on the Higgs branch which was described above for instance
does not get quantum corrected, see \cite{argyresseiberg}.  The quantum
corrections to the Coulomb branch are quite well known since the
work of Seiberg and Witten. Note that these arguments fail on the special points of the moduli space on which extra degrees of freedom become massless. Below, we will review the field theory
analysis of the quantum corrections to the Coulomb branch found
above.

\subsubsection{Saddle point analysis}
First we want to study the theory in a regime where the theory is
expanded around saddle points of the classical superpotential. We
therefore want to fix the eigenvalues of the adjoint such that
$W'$ vanishes for those values. Let us consider $W'(x) =
\prod_{i=1}^{n} (x-a_i)$. If $|a_i - a_j|
>> \Lambda, i \neq j$, then the saddle point analysis should give a good
description of the theory. The gauge group is broken to $U(N_i)$
subgroups centered at the $a$'s, where each subgroup has an
effective quadratic potential $\sim \mu \Phi^2$ for the adjoint
scalar of this subgroup, and the masses of the hypermultiplets get
shifted by $a_i$. We can therefore reduce the analysis to the
analysis of a $\mathcal{N} = 1$ theory perturbed by a quadratic
potential with a certain number of massless  fundamentals which can
be combined into an analysis for the general superpotential. The
vacuum structure now depends on the number of massless flavors
charged under this subgroup, $M_i$, since this number determines
the effective low energy degrees of freedom of the theory. These can be
determined, for instance, by analysis of the anomaly consistency
conditions (see e.g. \cite{seibergduality, shifman}). We will
analyze the system below mostly in the limit of vanishing masses and a
large but finite $\mu$.

\subsubsection{$M_i < N_i$}
The mass of the effective potential is very large (but finite), so
the adjoint scalars can be integrated out of the superpotential
(\ref{N2superpotential}) to give an effective theory for the meson
field $M_{i}^j = Q^a_i Q_a^j$. Now add to this lagrangian the
known $N=1$ non-perturbative contribution of the
Affleck-Dine-Seiberg \cite{affleckdineseiberg} superpotential to
arrive at the potential
\begin{equation}
{\mathcal{W}} = - \frac{1}{2 \mu} \tr{M^2} + (N_i - M_i)
\Lambda_i^{\frac{3 N_i - M_i}{N_i - M_i}} \det(M)^{\frac{-1}{N_i-M_i}}.
\label{potentialnflessnc}
\end{equation}
This potential is exact by the usual combination of holomorphy and symmetry arguments, supplemented by knowledge of weak coupling limits and analyticity, see e.g. \cite{horioogurioz}. 
The $\mathcal{N}=1$ scale $\Lambda_i$ can be related to the
$\mathcal{N} = 2$ scale by scale matching, $\Lambda_i^{3 N_i -
M_i} = \mu^N_i \Lambda^{2 N_i - M_i}$. We can now analyze the
minima of the above potential by using the technique outlined in
\cite{horioogurioz}. We find that the meson matrix can be
diagonalized by flavor rotations to a matrix with two different
entries. Let the number of entries of one type, say $\alpha_1$ be
given by an integer $r \leq [\frac{M_i}{2}]$. In general we will
have $2 N_i - M_i$ different solutions for every $\alpha_1$,
leading to $2 N_i - M_i$ different vacua at every $r$. If
$r=\frac{M_i}{2}$ then we have $N_i - M_i/2$ vacua. 

Bare quark masses can be added to (\ref{potentialnflessnc}), which generically break the full flavour symmetry group. For a given $r$ branch with massless quarks there are $\binom{M_i}{r}$ inequivalent ways of deforming this by giving masses to the quarks. The vacua obtained as minima of potentials of the form (\ref{potentialnflessnc}) will be referred to as 'electric' vacua.

\subsubsection{$M_i = N_i$}
In addition to the meson field degrees of (gauge invariant)
freedom this case also has a baryonic degree of freedom. The
effective $\mathcal{N} = 1$ theory is described by a linear sigma
model for the meson and baryons. Adding to this the potential
inherited from the $\mathcal{N} = 2$ theory we get an effective
potential
\begin{equation}
{\mathcal{W}} =  - \frac{1}{2 \mu} \tr{M^2} + K (\det(M) - B
\tilde{B} - \Lambda_1^{2 N_i}) .
\label{potentialnfequalnc}
\end{equation}
The vacua of this theory include the same type of vacua as for
$N_i < M_i$ when $B=\tilde{B}=0$, so we can analytically continue
the above counting results to this case. In addition to these,
there is one vacuum with non-zero expectation value for the
baryons $B,\tilde{B}$ .

\subsubsection{$M_i > N_i$}
This case is fundamentally different from the previous one: the
degrees of freedom in the low energy effective action are the
Seiberg dual \cite{seibergduality} magnetic quarks and gluons for
a gauge theory with dual gauge group $U(\hat{N_i} \equiv M_i -
N_i)$ and $M_i$ flavors. The effective superpotential for these
degrees of freedom is given by
\begin{equation}
{\mathcal{W}} _d = \tilde{q}_i M^i_j q^j - \frac{1}{2 \mu} \tr{M^2}
+ \left((N_i - M_i) \Lambda^{\frac{2 N_i - M_i}{N_i - M_i}} \cdot
\det(M)^{\frac{-1}{N_i-M_i}} \right),
\label{eq:seibergdualquarkpotential}\end{equation}
where the part between parenthesis is only added when the rank of
the meson matrix is $M_i$, since in this case the IR theory is a
pure glue theory. This superpotential has two branches of
solutions, depending on whether the meson matrix is degenerate or
not (compare the classical analysis for the Higgs branches). If it
is not degenerate, then the dual quarks can be integrated out and
the analysis is the same as in the case $M_i < N_i$. In
particular, we get the same number and degeneration pattern of
vacua as we obtained there. We will refer to this case as
'electric' vacua. If the meson matrix \emph{is} degenerate, the
solutions are classified by an integer $\hat{r}$, just as in the
previous analysis, but which now appears for the dual quarks and
mesons. We will refer to these vacua as 'magnetic'. The difference
in the situation in the electric case is the coupling to the meson
matrix. This coupling gives a mass to all quarks. In the IR we are
therefore left with a classical gauge group $U(\hat{N}_c -
\hat{r})$ with no massless flavors which gives $\hat{N}_c
-\hat{r}$ vacua with a surviving $U(1)$. In addition to these
two branches we also have a baryonic branch, just as in the
electric case.

\begin{table}
\caption{Number of extra vacua for $\cal N$ $= 2$ softly broken to
$\cal N$ $= 1$ by a quadratic superpotential. Here $\hat{N}_i \equiv N_c - N_f$.}
\label{table:phases}
\begin{center}
\begin{tabular}{lll}
case & \# vacua & labeled by \\
\hline
all $M_i$    &   $\binom{M_i}{r} ( 2 N_i - M_i)$       &   $r = 0,\ldots ,[M_i]/2]$ \\
        &   $\binom{M_i}{r} (N_i - \frac{M_i}{2})$    &   $r = M_i/2$\\
$M_i \geq N_i$ &   $1$                     &   $r = N_i$\\
$M_i > N_i$ &   $\binom{M_i}{\hat{r}} (\hat{N_i} - \hat{r})$ &    $\hat{r} = 0, \ldots, \hat{N_i}$ \\
        &   $1$                     &   $\hat{r} = \hat{N_i}$ \\
\end{tabular}
\end{center}
\end{table}

The weak coupling analysis now yields the table
\ref{table:phases}, where we have only listed phases which appear
\emph{in addition} to the ones already found in the cases listed
above it. The binomial factors are included to facilitate comparison to the results further in the paper. Note that the binomial factors are included and we
stress that this is the result for zero effective quark mass. In
this table we have abused notation slightly by labeling the
baryonic branches by the integer $r$ or $\hat{r}$.

\subsubsection{Strong coupling analysis}
\label{subsub:quantumu}

The final step consists of analyzing the Coulomb branch dynamics
encoded in the Seiberg-Witten curve of the gauge theory and especially on its singular points. We take the Seiberg-Witten differential $\lambda_{sw}$ from
\cite{dhokerkricheverphong96},
\begin{equation}
\lambda_{sw}(z) = \frac{z}{y} \left(P'(z) - \frac{1}{2} (P(z) - y(z))
\frac{Q'(z)}{Q(z)}  \right) dz = z\, \mathrm{d}\! \ln \left(y + P(z)\right),
\end{equation}
with  
\begin{equation}
y^2 = P(z)^2 - 4 \Lambda^{2 N_c - N_f} \left(z+m_i \right)^{N_f} \equiv P^2 - 4 Q 
\end{equation}
the Seiberg-Witten curve. As noted in \cite{dhokerkricheverphong96}, there are several proposals for the exact form of the Seiberg-Witten curve. Specifically, there can be a term of form $\Lambda^{2 N_c - N_f}$ times a polynomial of order $N_f-N_c$ if $N_f \geq N_c$ in the polynomial $P$. This corresponds to an inherent ambiguity in the definition of the quantum operators which reduce to $\frac{1}{k} \tr \Phi^k$ in the classical limit for $k \geq 2 N_c-N_f$. In general there are several possibilities. Specifying a definition of these operators in the quantum theory fixes the ambiguous polynomial, see e.g. \cite{schnitzer}. In this paper the natural definition by the gauge theory resolvent is used,
\begin{equation}
\frac{1}{k} \tr <\!\Phi^k\!>  =  \frac{1}{k} \frac{1}{2 \pi i} \oint_{\infty}dz z^{k-1},
\lambda_{sw}, \label{eq:gaugetheoryresolvent}\end{equation}
where the loop runs counterclockwise. Note that the Seiberg-Witten form also contains one power of $z$. 

Turning on a superpotential lifts most of the moduli space of the parent $\mathcal{N} = 2$
theory; the only points not lifted are those at which new degrees
of freedom (dyons) become massless or which are fixed by
non-zero expectation values of the quarks. In this paper we do
not consider theories in which mutually non-local dyons become
massless and we take all quarks to have the same mass $m$. Theories of the sort we are studying in this paper have been studied by various other methods in the literature, for instance by using geometric engineering \cite{ookouchi}, Konishi anomaly techniques \cite{cachazoseibergwitten}, exact potentials \cite{deboeroz}, or by using a brane setup \cite{deboeroz}. The upshot is that the gauge theory vacua are described by a factorized Seiberg-Witten curve of the form
\begin{eqnarray}
P(x)^2 - 4 Q(x) & =& (x+m)^{2 r}  H_{N_c-r-n+l}^2(x) F_{2n-2l}(x) \nonumber \\
(W'^{(c)}_n(x) + f_{n+1}(x)) &=&  B_l^2(x) F_{2n-2l}(x),
\label{eq:curvfact}\end{eqnarray} 
where $r$ labels the Higgs branches as before, $H$ is a polynomial in $x$ of order $N_c-r-n+l$, $B_l$ is a polynomial of degree $l$ and $W'^{(c)}_n(x)$ is the derivative of the classical superpotential.

In \cite{balastonaqvi} an extra contribution for $n \geq 2 N_c - N_f$ to the non-singular part of the Seiberg-Witten curve was reported by revisiting the field theory analysis in \cite{deboeroz}. The quantum operators used in that derivation however are \emph{not} the quantum operators defined through the resolvent. In appendix \ref{ap:curvfact} it is shown that if these quantum operators are defined through the gauge theory resolvent the extra contribution in the calculation disappears. 

The baryonic Higgs branch is much more easy, since the gauge theory is in this case fully Higgsed. The only way this can be reflected in the Seiberg-Witten curve is the case when the whole curve factorizes, $y^2=H^2$.

%\begin{equation}
%\frac{1}{k} \tr <\!\Phi^k\!> = \frac{-1}{2 \pi i} \oint_{\infty}dz z^{k-1} \ln \left(y + P(z)\right) + \frac{1}{2 \pi i k} \left[z^{k} \ln \left(y + P(z)\right) \right]_{z= R e^{-i \pi}}^{z= R e^{i \pi}}
%\end{equation}
In the following we will repeatedly use a partially integrated form of expression (\ref{eq:gaugetheoryresolvent}). Partial integration will give a boundary term in the loop integral because the logarithm has a branch cut which we choose to be on the negative real axis. However, this boundary term mostly cancels in calculations (as it should) since we will only calculate the pole at infinity. Here the function $\ln \left(y + P(z)\right) \sim \ln \left(z^{N} + l.o.\right) = \ln(z^N) + \ln(1 + \mathcal{O}(1/z))$. The last part is a nice holomorphic function in a neighborhood of $z=\infty$. Now the boundary term cancels against the term coming from $\oint z^{k-1} \ln z^{N}$. The only exception to this canceling mechanism are terms
\begin{equation}
\frac{1}{2 \pi i} \oint d(\log(z^N)) = N.
\label{eq:exception}\end{equation}
These terms we will however not need in this paper since they correspond to the trace of the identity operator. We can therefore write
\begin{equation}
\frac{1}{k} \tr <\!\Phi^k\!> = \frac{-1}{2 \pi i} \oint_{\infty}dz z^{k-1} \ln \left(\frac{1}{z^{N_c}}(y(z) + P(z))\right).
\label{eq:partintresol}\end{equation}

\subsection{Compactifications on $\mathbb{R}^3 \times S^1$}

Above we reviewed the vacuum structure of four-dimensional ${\cal
N}=1$ gauge theories. Once we compactify on a circle, the vacuum
structure remains the same, as explained in the introduction.
However, the natural moduli of the theory are different. The
adjoint scalar of the four-dimensional theory remains an complex
adjoint scalar on $\mathbb{R}^3 \times S^1$, but the 4d gauge field
yields two additional scalars, one being the component of the 4d
gauge field in the direction of the $S^1$, the other being the
dual of the 3d gauge field. Both scalars are compact. At a generic
point on the moduli space of the ${\cal N}=2$ theory compactified
on $\mathbb{R}^3 \times S^1$, the gauge group is broken to
$U(1)^N$, and the moduli space is therefore $4N$-dimensional. In
the analysis using the spin chain below, we will encounter complex
variables $p_i$ and $q_i$. In the classical limit, they can be
roughly thought of as the vacuum expectation values of the adjoint
scalar $\Phi$ and of the two additional scalars respectively. The
precise relation of these variables to the quantum operators in
the theory is something we will not discuss in detail, but will certainly suffer from an  ambiguity similar to the one discussed in the previous subsection. In addition, from the integrable system point of view any set of variables $p'_i,q'_i$ that is related to the
original variables $p_i,q_i$ by a canonical transformation is perfectly acceptable. We will see this explicitly in the next sections. Again, which of the canonical variables have the simplest interpretation in terms of the microscopical definition of the gauge theory is something which is outside the scope of the
present paper.

\section{The classical $SL(2, \mathbb{C})$ spin chain and its degenerations} \label{sec:spinchain}
In this section we review the classical inhomogeneous twisted
periodic $SL(2)$ spin chain. We follow mainly \cite{gorskygukov}.
The curve with fundamental hypermultiplets is a generalization of
the curve without, so the integrable system lurking behind should
reduce in some limit to the periodic Toda chain. Since the
spin-chain will be formulated in terms of $2$ by $2$ matrices, the periodic Toda system will first be written in terms of such matrices.

\subsection{Periodic Toda in terms of 2x2 matrices}
Periodic Toda (see e.g. \cite{mumfordmoerbeke}) can be defined by
an infinite-dimensional Lax matrix which commutes with the
shift-by-N operator (is N-periodic). We can write this infinite
dimensional matrix acting on an eigenvector as a recursion
relation,
\begin{equation}
\Lambda^2 e^{q_{n} - q_{n-1}} f_{n-1} + \phi_{n} f_n + f_{n+1} = \lambda f_n,
\end{equation} which can be written as a matrix equation (in just
'site-local' variables for the matrix),
\begin{equation}
\left( \begin{array}{c} f_{n+1} \\ e^{q_{n}} f_n \end{array}\right) =
 \left( \begin{array}{cc} \lambda - \phi_n & - \Lambda^2 e^{- q_{n}} \\ e^{q_{n}} & 0
 \end{array}\right)  \left( \begin{array}{c} f_{n} \\
  e^{q_{n-1}}f_{n-1} \end{array} \right). \label{todachain}
\end{equation} Applying N matrices of this type will lead to the
shift-by-N Toda transfer matrix $T_{Toda}$:
\begin{equation}
T_{Toda} = \prod_{i=N}^1 \left( \begin{array}{cc} \lambda -
\phi_i & - \Lambda^2 e^{- q_{i}} \\ e^{q_{i}} & 0  \end{array}\right).
\end{equation} Since our original infinite dimensional matrix was
periodic with period $N$, we should focus on eigenvalues of the
transfer matrix. These eigenvalues are given by the roots of the
characteristic equation of this $2$ by $2$ matrix which is
\begin{equation}
\det (T_{Toda} - w) = w^2 - \tr T_{Toda} w + \det (T_{Toda})=0.
\end{equation} By defining $y = 2 w - \tr T_{Toda}$ and using $\det
(T_{Toda})=\Lambda^{2 N}$ we get the familiar form of the spectral
curve,
\begin{equation}
y^2 = (\tr T_{Toda})^2 - 4 \Lambda^{2 N}.
\end{equation}

\subsection{\xxx}
The above formulation of the periodic Toda chain can be used to
search for generalizations. Let us consider the following matrices
(confusingly also called Lax matrices),
\begin{equation}
L_j(\lambda) =  \lambda - \lambda_j + \sum_{k=1}^{3} S_k^{(j)} \sigma^k,
\end{equation}
\begin{equation}
L_j =  \left( \begin{array}{cc} \lambda - \lambda_j + S_1^j& S_2^j - i S_3^j \\ S_2^j +
i S_3^j & \lambda - \lambda_j - S_1^j  \end{array}\right),
\end{equation} where $\lambda_i$ are constants (trivial Casimirs) known
as impurities and the $\sigma^j$ are the Pauli matrices. We define
$\sigma^0$ to be the identity matrix. For this system to be
integrable the Lax matrices should satisfy quadratic r-matrix
relations (see e.g. \cite{babelon}, chapter 2),
\begin{equation}
\{L_i(\lambda) \stackrel{\otimes}{,} L_j(\lambda')\} = \delta_{ij}
 \left[r(\lambda-\lambda'),L_i(\lambda)\otimes L_j(\lambda')\right] \label{integracond},
\end{equation}  where we choose a rational r-matrix
$\frac{1}{2(\lambda' - \lambda)} \sum_{a=0}^{3} \sigma^a \otimes
\sigma^a$. This r-matrix is proportional to the interchange
operator $P: A \otimes B \rightarrow B \otimes A$ and the r-matrix
relation is invariant under multiplication on the left or right of
both Lax matrices with a constant matrix.  With this r-matrix the
$S_j$ should satisfy a classical $SL(2, \mathbb{C})$ Poisson
algebra,
\begin{equation}\label{poissonalg}
\{S_i^{a},S_j^{b} \} =  i \delta_{ab}  \epsilon_{ijk} S_k^{a}.
\end{equation}
Now define the transfer matrix and the associated spectral curve
in analogy to the Toda case,
\begin{equation}
T(\lambda) = \prod_{j=N}^1 L_j(\lambda)
\end{equation}
\begin{equation}
\det (T(\lambda) -w) = w^2 - \tr T w + \det (T) \equiv w^2 - P_N(\lambda) w + Q_{2N}(\lambda) =0,
\end{equation} where the subscripts denote the degree of the
polynomials. Defining $y = 2 w - \tr T_{Toda}$ we obtain
\begin{equation}
y^2 = (P_N)^2 - 4 Q_{2 N}.
\end{equation}
An important step in analyzing integrable systems is obtaining its conserved quantities. The spin chain variables have a natural quadratic Casimir, the spin length $(S^a)^2$, which commutes with all $S^b_i$. If we fix the values of the Casimirs, there remains a
Poisson manifold of dimension $2N$. The remaining variables can be
expressed in terms of $N$ coordinates and momenta (see for
instance appendix \ref{sec:heisvar}) with the usual Poisson bracket.
To completely solve the integrable system we therefore need
determine $N$ conserved quantities. Observe that products of Lax
matrices at different sites obey the same r-matrix relations, e.g.
$\{L_1(\lambda) L_2(\lambda) \stackrel{\otimes}{,} L_1(\lambda')
L_2(\lambda')\} = [r(\lambda-\lambda'), L_1(\lambda)
L_2(\lambda)\otimes L_1(\lambda') L_2(\lambda')]$. Taking the
trace on both factors in the direct product now gives a conserved
quantity of the spin chain. A convenient basis can be obtained by
studying $\{T(\lambda), T(\lambda')\}$, which gives the
coefficients of the trace of the transfer matrix as conserved
quantities.

Comparing with the Seiberg-Witten curve for $U(N_c)$ with $N_f = 2
N_c$ flavours in e.g.  \cite{argyresshapere} shows that the $Q$
term should be (up to the modular functions of the conformal gauge
coupling factor $e^{i \tau}$, which we will discuss momentarily)
$\prod_{i=1}^{2 N_c}(\lambda + m_i)$, where $m_i$ are the masses
of the fundamental hypermultiplets\footnote{strictly speaking the
$m_i$ as used here are modular functions themselves. If we take a
limit to the case of $N_f<2 N_c$, the corresponding limit on the
scale of the problem will reduce the $m(\tau)$ to simple constant
masses.}. In the \xxx chain the polynomial $Q_{2N}$ is,
\begin{equation}
Q_{2N}(\lambda) = \prod_{i=1}^N ((\lambda - \lambda_i)^2 - K_i^2),
\end{equation} where $K_i^2$ is the obvious quadratic Casimir of the
Poisson algebra (\ref{poissonalg}) at site $i$. We therefore obtain 2 masses from each site which are furthermore equal if $K_i$ is zero. Note that the full
spectral curve of the spin chain only contains Casimirs of the
algebra and is therefore itself a conserved quantity. 

As a final step in identifying gauge theory quantities, the modular functions have to be inserted in front of the $Q$, or for the more general
($N_f<2N_c$) case introduce a scale into the problem. It has already been noted that the integrability condition (\ref{integracond}) is invariant under multiplication of the Lax matrices by a constant matrix. We can therefore take generalized boundary
conditions without affecting integrability, or in other words take
a modified transfer matrix $\tilde{T} = C T$, where we take $C$ to be
\begin{equation}
C= \left( \begin{array}{cc} 1 & 1 \\ \frac{-1}{4} (1-g^2) & 0  \end{array}\right).
\end{equation}
In the case where $N_f=2 N_c$, $g$ is the modular function
$\frac{\theta^4_2 + \theta^4_1}{\theta^4_2 - \theta_1^4}$. To
really fit the curve in \cite{argyresshapere} $\lambda$ should be shifted by $g \mu$ with $\mu = \frac{1}{2 N_c} \sum m_i$ and the masses should be redefined in the spin chain to be $\mu_i =  m_i - \mu$. However, since we will be reducing to non-finite gauge theories in
a moment we will not do that here. In any case, in most cases considered in
this paper the gauge theory scale will be introduced in another way than by the above $C$ matrix, see subsection \ref{subsub:locallimits}.

\subsection{Degenerations}
In gauge theory the degeneration to lesser amounts of flavors is
performed by integrating out massive degrees of freedom. On the
level of the Seiberg-Witten curve this translates into taking
several masses to infinity while keeping the product of these
masses with the gauge theory effective scale fixed to a new
effective scale. If our integrable models describe the gauge
theory correctly, there should be a limit in the integrable system
which reproduces this procedure. We will discuss how to take
limits on the spin chain to integrate out masses. For one site and
in different variables this was discussed in \cite{gorskygukov}.

The $SL(2,\mathbb C)$ algebra at each site has a Casimir
which we will want to keep fixed in this paper since it relates to the
masses. Also the connection to the Toda chain is not clear in the
variables $S_i$, so we will re-express the $SL(2,\mathbb C)$
algebra in new variables $q,p,K$ for each site. The
Heisenberg-like variables $q,p$ obey the standard Poisson algebra
$\{q_i,p_j\}= \delta_{ij}$. An additional reason to switch to
Heisenberg variables is that the 3d superpotential we will
consider has known non-perturbative contributions in the pure glue
sector of the form $\sum e^{q_{j+1} -q_j}$ and these appear
naturally in Heisenberg like variables. Hence, these new
variables are much closer to the naive gauge theory degrees of
freedom. After some analysis (see the appendix
(\ref{sec:heisvar})) we arrive at the Lax matrix:
\begin{equation}
L_j =  \left( \begin{array}{cc} \lambda - p_j &  (m_{2j} + p_j)e^{-q_j}
\\- (m_{2j-1} + p_j)e^{q_j} & \lambda +  (m_{2j} +
 m_{2 j -1}) + p_j  \end{array}\right), \label{xxxlaxheis}
\end{equation}
where one of three possible forms is chosen,
namely the one which is the most convenient one for taking limits. The other two forms are connected to this one by a (possibly singular) canonical transformation. In this form the inhomogeneities $\lambda_j$ and the Poisson-algebra Casimirs $K_j$ have been expressed in terms of the masses. We want to study degenerations to the Lax matrix of the Toda chain(\ref{todachain}) which can be done by taking certain scaling
limits. Immediately a problem presents itself: if we introduce the
modular functions through a $C$ matrix as above, then not all
masses are multiplied by the coupling function in general which
prevents us from taking masses to infinity directly. The origin of this
problem is the site-locality of the masses at sites and the site
global form of the coupling matrix. Since the masses are site-local, we will study a local limiting procedure below. However, this global versus local problem is the source of the existence of the different coordinate patches we will encounter later.

\subsubsection{Local limits}\label{subsub:locallimits}
As noted in the previous subsection, the integrable structure of
the chain is invariant under multiplication by a constant matrix.
Multiply (\ref{xxxlaxheis}) by constant matrices  $A_J = \left(
\begin{array}{cc} 1 & 0 \\ 0 & \alpha_j \end{array}\right)$ and
$B_J = \left(\begin{array}{cc} 1 & 0 \\ 0 &\beta_j
\end{array}\right)$ on the right and left respectively to obtain
\begin{equation}
L_j =  \left( \begin{array}{cc} \lambda - p_j &  \alpha_j(m_{2j} +
p_j) e^{-q_j} \\ - \beta_j(m_{2j-1} + p_j) e^{q_j} & (\lambda +
(m_{2j} + m_{2 j -1}) + p_j )\alpha_j \beta_j \end{array}\right),
\label{xxxlaxheis2}
\end{equation}
At every site there are four different limits one can take:

\begin{enumerate}
\item Take both $\alpha_j$ and $\beta_j$ to zero,
keeping $\beta_j m_{2j-1}$ and $\alpha_j m_{2j}$
fixed (at $\Lambda$). We will call this limit $(0)$
\item Take $\alpha_j$ to zero and $\beta_j$ to one,
keeping  $\alpha_j m_{2j}$ fixed (at $\Lambda$).
We will call this limit $(1^u)$.
\item Take $\beta_j$ to zero and $\alpha_j$ to one,
keeping  $\beta_j m_{2j -1}$ fixed (at $\Lambda$).
We will call this limit $(1^d)$.
\item The trivial limit $\alpha=1$ and $\beta=1$,
which we will call $(2)$
\end{enumerate}

Note that the first limit reproduces the Toda chain 2 by 2
matrices. The gauge theory scale is introduced in the above
by dimensional analysis and by comparing to the Toda chain.
In principle we could also take an entirely different set of
limits where in the end all the top left components of the Lax
matrix would be put to zero when taking the Toda limit.  However,
it seems inconsistent to take such a limit on one of the Lax
matrices and then take the above limit $(0)$ on the next one: this
procedure would for instance not reproduce a Seiberg-Witten curve of $U(N_c)$ gauge theory without flavors in the case where all masses are integrated out. Since taking
the above described limit or the alternative limit at \emph{all}
matrices will give the same spectral curve we will only consider
the ones in the list above.

We will use the following notation for limits: the total quark
content of the chain will be given by a vector $M := (e_{N_c},
e_{N_c-1}, \ldots, e_1 )$, where $e_i$ indicates the limit taken
on a site.

\section{The proposal}
\label{sec:proposal}
In this section we present our proposal for the exact superpotential of $\mathcal{N} =
2$ $U(N_c)$ gauge theory with $N_f$ hypermultiplets in the
fundamental representation on the manifold $\mathbb{R}^3 \times
S^1$, which is broken to $\mathcal{N} = 1$ by adding a
(polynomial) superpotential $W$ for the adjoint superfield $\Phi$.

\subsection{Motivation}
As explained in the introduction, in order to write down the
quantum superpotential we merely have to find good holomorphic
variables on the moduli space. The integrable system and its
degeneration described in the previous subsection has all the
required properties and therefore provides us with variables that
are valid on at least an open subset of the moduli space.
Therefore, our proposal simply boils down to using the conserved
charges of the integrable system to build the quantum
superpotential. There is a subtlety regarding exactly which
conserved quantity we should associate to the operator ${\rm
tr}(\Phi^k)$. As discussed in section \ref{subsub:quantumu} we will use to gauge theory resolvent to give a precise quantum meaning to ${\rm tr}(\Phi^k)$ and to define it
unambiguously in terms of the conserved charges of the integrable
system. This quantum definition is also the one for which the factorization of the Seiberg-Witten curve takes on the simple universal form of equation (18).
\subsection{The superpotential}
The form of the superpotential in 4d for the softly broken
$\mathcal{N} = 2$ theory is
\begin{equation}
W = W(\Phi) + \sqrt{2} \tilde{Q}^i_a m_i^j Q_j^a + \sqrt{2} \tilde{Q}^i_a \Phi^a_b Q_i^b.
\end{equation}
In the pure glue case it was natural to replace the moduli space
coordinates in the superpotential by (a specific basis of) the
conserved quantities (Hamiltonian functions) in the integrable
system, i.e. traces of powers of the Lax matrix. In the present
case that would translate to replacing the moduli space
coordinates by certain hamiltonians of the \xxx chain. In the
direct application of this idea we then run into the problem of
what to do with the $Q$'s: since there are no dynamical variables
in the chain which correspond naturally to the fermionic quark
fields or the composite meson field variables. This is highly
reminiscent of the lack of natural glueball field variables in the
periodic Toda chain. Note that, just as the on-shell glueball
fields can be calculated from the factorization of the curve, on-shell we can recover the meson vevs \cite{cachazoseibergwitten} as well. Imagine integrating out the
quarks, irrespective of whether this is allowed. We are then left
with an effective superpotential of the same degree as the
original one. However, in order to integrate out the quarks we need to choose a
particular vacuum around which to expand. One would therefore expect to find only one branch, with fixed $r$, of the full vacuum structure. For an attempt to calculate an integrable system potential in a similar way see appendix \ref{sec:anobservation}.

In the main body of this paper we will follow a more naive
approach by taking $W(\Phi)$ and replacing the gauge theory
Casimirs directly by a certain basis of the Hamiltonians of the
spin chain. Perhaps not surprisingly, we will find that the
equilibria of the resulting superpotential generically describe
several $r$ branches at the same time. To make the proposal concrete we have to specify how to translate the gauge theory traces in the superpotential to an expression in the conserved charges of the \xxx chain. As stated above, and motivated by the strong
coupling analysis in the gauge theory, we take the approach of
using the gauge theory resolvent to define the operators ${\rm Tr}<\!\Phi^k\!>$ in the integrable system. This proposal reduces to the pure Toda chain one \cite{firstpaper} in the Toda limit (on all the sites).

In the following several vacua of this proposed superpotential will be studied. We will find that there can be different solutions to the equations of motion which yield the same Seiberg-Witten curve. These solutions therefore describe the same vacuum from the
four dimensional point of view. However, on $\mathbb{R}^3 \times
S^1$ they obviously correspond to inequivalent vacua.

\section{Addition and multiplication maps}
\label{sec:admultmaps}
On field theory grounds some vacua of gauge theories with different matter content and/or different rank of the gauge group should be connected through the so-called addition and multiplication maps (see for instance \cite{balastonaqvi}). In this section we show that these maps both have a natural counterpart in the integrable system. We will use that the e.o.m. derived from a polynomial superpotential can be written as,
\begin{equation}
\frac{\delta W}{\delta o_i} = \frac{-1}{2 \pi i} \oint_{\infty} W'(x) \frac{\frac{\delta P}{\delta o_i}}{\sqrt{P^2 - 4 Q}}.
\label{eq:eqofmotiongen}\end{equation}
Here $o_i$ is either $p_i$ or $q_i$. Hence if the derivative of the transfer matrix with respect to the coordinates vanishes the equations of motion are satisfied.

%\begin{equation}
%\frac{\delta W}{\delta o_i} = \sum_{k=0}^{n+1} g_k \frac{\delta u_k}{\delta o_i} =  \sum_{j=0}^{n+1} \sum_{k=0}^{n+1} g_k \frac{\delta u_k}{\delta s_j}\frac{\delta s_j}{\delta o_i}.
%\end{equation}
%Here $o_i$ is either $p_i$ or $q_i$. Hence the e.o.m are equivalent to the vanishing of the derivative of the transfer matrix with respect to the coordinates.

\subsection{Addition map}
Suppose we have a solution of the spin chain e.o.m derived for a fixed superpotential $W = \sum_{k=0}^{n+1} \frac{g_k}{k} \tr(\Phi^k) = \sum u_k g_k$ in a $N_c, N_f$ parent gauge theory. Using this it must then be shown that (some of the) solutions to the e.o.m of spin chain corresponding to a daughter gauge theory with the same superpotential and $\tilde{N}_c = N_c+1, \tilde{N}_f=N_f +2$ can be generated. Likewise, there must be a solution in the $\tilde{N}_c, \tilde{N}_f$ which can be projected down to the parent theory. In the field theory analysis \cite{balastonaqvi} this map is known as the addition map and below we show how it has a natural interpretation in our proposal. 

Start with the transfer matrix of the $N_f,N_c$ theory and add one diagonal Lax matrix $(x+\tilde{m})\mathbb{I}_{2 \times 2}$ to it, which can be obtained by adding a spin chain Lax matrix with 2 masses $\tilde{m}$ and setting $p \rightarrow -\tilde{m}$. Then it easily seen that the spectral curve of the resulting transfer matrix has the form $(x+\tilde{m})^2(P_N^2 - 4 Q)$ appropriate for a $\tilde{N}_c = N_c+1, \tilde{N}_f=N_f +2$ theory, with $2$ flavours of mass $\tilde{m}$ on the Higgs branch. It then must be shown that the e.o.m. of the two theories have a simultaneous solution which has $p \rightarrow -\tilde{m}$, so that a solution in the parent theory can be lifted to the daughter. The generating function for the e.o.m. in the $\tilde{N}_c$ theory is obtained by adding a matrix at site $N+1$ to a $U(N_c)$ transfer matrix, 
\begin{equation}
\tr T_{\tilde{N}_c}(x) = \tr \left( \left( \begin{array}{cc} x - p_{\tilde{N}_c} &  (\tilde{m} +
p_{\tilde{N}_c}) e^{-q_{\tilde{N}_c}} \\ - (\tilde{m} + p_{\tilde{N}_c}) e^{q_{\tilde{N}_c}} & x +
2 \tilde{m} + p_{\tilde{N}_c}  \end{array}\right) \left(\begin{array}{cc} t_{11}& t_{12}\\ t_{21} & t_{22} \end{array}\right) \right),
\end{equation}
\begin{eqnarray}
\tr T_{\tilde{N}_c}(x) &=&  (x - p_{\tilde{N}_c}) t_{11} +  (\tilde{m} + 
p_{\tilde{N}_c}) e^{-q_{\tilde{N}_c}} t_{21}  - \nonumber \\
& &\qquad \qquad  (\tilde{m} + p_{\tilde{N}_c}) e^{q_{\tilde{N}_c}} t_{12} + (x + 2 \tilde{m} + p_{\tilde{N}_c} ) t_{22}.
\end{eqnarray}
It is easy to see that the e.o.m. for $o_i \in {p_i, q_i}$ $i \neq N_c +1$ simply project down to the $U(N_c)$ theory when $p_{\tilde{N}_c} = -\tilde{m}$, since only the $t$'s depend on these coordinates. In this case the e.o.m. can be derived from  
\begin{equation}
\frac{\delta T_{\tilde{N}_c}}{\delta o_i} = (x+m) \frac{\delta T_{N_c}}{\delta o_i} .
\end{equation} 
The equation of motion for $q_{\tilde{N}_c}$ is also solved by this value for $p_{\tilde{N}_c}$. The equation obtained by varying with respect to $p_{\tilde{N}_c}$ is quadratic in $e^{q_{\tilde{N}_c}}$ on this solution since in that case,
\begin{equation}
\tr T \sim t_{22} -t_{11}  - e^{-q_{\tilde{N}_c}} t_{21} - e^{q_{\tilde{N}_c}} t_{12},
\end{equation}
and the Seiberg-Witten curve is independent of $e^{q_{\tilde{N}_c}}$ on the solution. The resulting equations can be solved for $e^{q_{\tilde{N}_c}}$ which yields two solutions. The solutions for $p_{\tilde{N}_c}$ and $q_{\tilde{N}_c}$  map two solutions to the e.o.m. of the daughter theory to the parent. These two solutions for $q_{\tilde{N}_c}$ correspond to the same Seiberg-Witten curve in the parent theory. Vice-versa, one solution to the parent theory e.o.m. can be lifted to two solutions in the daughter with equivalent Seiberg-Witten curve. It remains to be shown that the e.o.m. projected down to the parent theory can be derived from the same superpotential in both theories, but this follows trivially from (\ref{eq:eqofmotiongen}). Note that on shell we have 
\begin{equation}
\tilde{u_k} = \frac{1}{k} (\tilde{m})^k + u_k.
\label{eq:41}\end{equation}

\subsection{Multiplication map}
The multiplication map can be derived in the $2 \times 2$ formalism just as it was for the pure glue gauge theory in the $N_c \times N_c$ form in \cite{firstpaper}. The derivation below is the extension of the integrable systems multiplication map to theories with fundamental matter. Consider a transfer matrix for a $U(t N_c)$ gauge theory with $t N_f $ quarks which is the product of t transfer matrices $T_{N_c}$ of the parent $U(N_c)$ theory. 

It needs to be shown that solutions to the e.o.m. for the $U(tN_c)$ can be projected solutions to the e.o.m. of the $U( N_c)$ theory by periodically identifying coordinates and momenta. The difference between the e.o.m. for $q_j$ and $q_{a t+j}$ can be studied in the $U(tN_c)$ theory, with $a$ an integer,
\begin{equation}
\left(\frac{\delta }{\delta q_j } - \frac{\delta}{\delta q_{at+j}} \right) P_{t N_C}.\label{eq:checkperiodic} 
\end{equation}
The transfer matrix can be written in a site-local form. Hence taking derivatives with respect to coordinates $q_j$ and $q_{a t+j}$ (or momenta) and periodically identifying with period $t$ will yield the same transfer matrix in both cases. Therefore the right hand side of equation (\ref{eq:checkperiodic}) vanishes and the e.o.m. of the $U(tN_c)$ theory can be projected down to $U(N_c)$ theory. 

%There these same e.o.m. can be derived by using the same classical superpotential since 
%\begin{equation} u_k^{tN_c} = t u^{N_c}_k. \end{equation} 

It remains to be shown that the equations of motion can be derived from the same superpotential and how the Seiberg-Witten curves are related. Below we will first show that the curve in the $U(tN)$ theory is related to the curve in the $U(N)$ theory as 
\begin{equation}
P_{tN_c} ^2 - 4 Q_{t N_c} = H_{2(t-1) N_c}^2 ( P^2_{N_c}(x) - 4 \det(T_{N_c})),
\end{equation}
with $H$ a polynomial of order $2(t-1) N_c$. In particular, if one curve factorizes as in equation ($\ref{eq:curvfact}$) for a particular solution in that theory, so does the multiplication map of that curve when evaluated on the solutions which have been generated by the multiplication map of that solution. 

The trace of the product of transfer matrices, $\tr T_{tN_c}$, only depends on the two eigenvalues of $T_{N_c}$, so the trace of the product of $t$ of them can be written as a function of the trace and the determinant of $T_{N_c}$. In fact, since for 2 by 2 matrices
\begin{equation}
\tr(T_{tN_c}) = \tr(T_{N_c}^t) = \tr(T_{N_c}) \tr(T_{N_c}^{t-1}) - \det(T_{N_c}) \tr({T_{N_c}}^{t-2}),
\end{equation}
we have
\begin{equation}
\frac{\tr(T_{N_c}^t)}{(2 \det(T_{N_c}))^{t/2}} = \frac{2 \tr(T_{N_c})}{2 \sqrt{\det(T_{N_c})}} \frac{\tr(T_{N_c}^{t-1})}{2 \det(T_{N_c})^{(t-1)/2}} - \frac{\tr(T_{N_c}^{t-2})}{2 (\det(T_{N_c}))^{(t-2)/2}},
\end{equation}
or defining $C_t = \frac{\tr(T_{N_c}^t)}{2 (\det(T_{N_c}))^{t/2}}$ and $x \equiv C_1 = \frac{\tr(T_{N_c})}{2 \sqrt{\det(T_{N_c})}}$
\begin{equation}
C_t(x)= 2 x C_{t-1}(x) - C_{t-2}(x),
\end{equation}
which is the Chebyshev recursion relation (for Chebyshev polynomials of the \emph{first} kind with the above initial conditions). Putting it all together we get,
\begin{equation}
P_{tN_c}(x) = 2 (\det(T_{N_c}))^{t/2} C_t\left(\frac{P_N(x)}{2 \sqrt{\det(T_{N_c})}}\right),
\end{equation}
as is expected on purely field theory grounds (see e.g. \cite{civ}) and which combines nicely with the determinant of $T_{tN_c}$ to give the correct factorization of the curve.
\begin{eqnarray}
P_{tN_c} ^2 - 4 Q_{t N_c}& =& 4 \det (T_{N_c})^{t} \left(C^2_t \left( \frac{P_{N_c}(x)}{2 \sqrt{\det(T_{N_c})}} \right) - 1 \right) \nonumber \\
 & = & \det (T_{N_c}))^{t-1} U_{t-1}^2 \left( \frac{P_{N_c}(x)}{2 \sqrt{\det(T_{N_c})}} \right) ( P^2_{N_c}(x) - 4 \det(T_{N_c})).
\end{eqnarray}

Note that the first part of the last equation can be written as the square of a \emph{polynomial}. This can be easily proven using the parity relations for the Chebyshev polynomials. Since the scale in the $U(tN_c)$ theory must be real, $\det(T_{N_c})$ must be a $t$-th root of unity and there are $t$ inequivalent ways of lifting a solution of the $U(N_c)$ theory to a solution of the $U(tN_c)$ gauge theory. 

Using the above analysis of the Seiberg-Witten curve, it can now be shown that the equations of motion of parent and daughter theory in the multiplication map are derived from the same superpotential. From equation (\ref{eq:eqofmotiongen}) we get that
\begin{equation}
\frac{\delta W}{\delta o_i} = \frac{-1}{2 \pi i} \oint_{\infty} W'(x) \frac{C'_t\left(\frac{P_N(x)}{2 \sqrt{\det(T_{N_c})}}\right) \frac{\delta P_{N}}{\delta o_i}}{ U_{t-1} \left( \frac{P_{N_c}(x)}{2 \sqrt{\det(T_{N_c})}} \right) \sqrt{P_N^2 - 4 Q_N}}.\end{equation}
The following relation can easily be proven using the generating functions for the Chebyshev polynomials,
\begin{equation}
C'_t(x) = t U_{t-1}(x),
\end{equation}
and it is then easy to see that the equations of motion in both parent and daughter theory are derived from the same superpotential.  Note that there is no freedom to choose limits in these versions of the  multiplication and addition maps, although when the above maps are combined with finite non-singular canonical transformations there certainly is such freedom. 

\section{Examples}
\label{sec:examples} 
This section contains several examples in which the vacua of several of our proposed superpotentials are calculated. Unless stated otherwise these examples are for a quadratic superpotential with $W'(-m) = 0$ and with equal mass fundamental hypermultiplets. 

\subsection{$N_c = 2$}
\subsubsection{$N_f=2$}
We will do this case in some detail, to illustrate the
(in)equivalence of the various limits that one can take on the
full \xxx chain. In the notation introduced in section~4, there
are three different and inequivalent limits that one can take,
namely $(2,0)$, $(1^u,1^u)$ and $(1^u,1^d)$. Below we list for
each of these three cases the value of $P_2(x)$, $Q_2(x)$,
$u_1={\rm Tr}(\Phi)$ and $u_2=\frac{1}{2} {\rm Tr}(\Phi^2)$. The
latter we compute using the gauge theory resolvent, so that it is
slightly different from the naive quantity that one would extract
from $P_2(x)$ (see equation (\ref{eq:difftrphisq})). For simplicity we also introduce the notation $y=e^{q_1-q_2}$. We find for the case $(2,0)$
\begin{eqnarray}
P_2(x) & = & (x-p_1)(x-p_2) - y \Lambda (p_1+m_1) - y^{-1} \Lambda
(p_1+m_2) \nonumber \\
Q_2(x) & = & \Lambda^2 (x+m_1)(x+m_2) \nonumber \\
u_1 & = & p_1+p_2 \nonumber \\
u_2 & = & \frac{1}{2} (p_1^2+p_2^2) + \Lambda^2 + \Lambda y
(p_1+m_1) + \Lambda y^{-1} (p_1+m_2),
\end{eqnarray}
for the case $(1^u,1^u)$
\begin{eqnarray}
P_2(x) & = & (x-p_1)(x-p_2) - y \Lambda (p_1+m_1) - y^{-1} \Lambda
(p_2+m_2)+\Lambda^2  \nonumber \\
Q_2(x) & = & \Lambda^2 (x+m_1)(x+m_2) \nonumber \\
u_1 & = & p_1+p_2 \nonumber \\
u_2 & = & \frac{1}{2} (p_1^2+p_2^2)  + \Lambda y (p_1+m_1) +
\Lambda y^{-1} (p_2+m_2),
\label{eq:upup}\end{eqnarray}
and finally for $(1^u,1^d)$
\begin{eqnarray}
P_2(x) & = & (x-p_1)(x-p_2) - y \Lambda^2 - y^{-1}
(p_1+m_1)(p_2+m_2)+\Lambda^2  \nonumber \\
Q_2(x) & = & \Lambda^2 (x+m_1)(x+m_2) \nonumber \\
u_1 & = & p_1+p_2 \nonumber \\
u_2 & = & \frac{1}{2} (p_1^2+p_2^2)  + \Lambda^2 y  +  y^{-1}
(p_1+m_1)(p_2+m_2).
\end{eqnarray}
As we expected, these three cases are related to each other by
canonical transformations. The second case is obtained from the
first one by the transformation
\be
p_1 \rightarrow p_1 -\Lambda y^{-1},\qquad p_2 \rightarrow
p_2+\Lambda y^{-1}
\ee
while the third case is obtained from the second one by the
transformation
\be
y\rightarrow \Lambda y/(p_1+m_1).
\ee
Notice that this last transformation is not well-defined
everywhere on the phase space. This illustrates the general
phenomenon that different degenerations can in principle describe
different open subsets of moduli space, though on overlaps they
are always related by a canonical transformation.

The vacua of a purely quadratic superpotential (proportional to
$u_2$) in the case of equal masses $m_1=m_2=m$ are easily found.
In the third case $(1^u,1^d)$, there are three types of solutions.
The first are the $r=0$ solutions
\be
p_1=p_2=\pm \Lambda , \quad y=-\frac{\Lambda\pm m}{\Lambda}
\ee
for which
\be
P_2^2(x)- 4 Q_2(x) = (x\mp 2\Lambda)^2 (x^2 \pm 4 m \Lambda + 4
\Lambda^2).
\ee
This follows the general factorization pattern summarized in (\ref{eq:curvfact}). Secondly, we have the $r=1$
solutions
\be p_1= \frac{-m\pm \sqrt{m^2-4\Lambda^2}}{2},\quad
p_2= \frac{-m\mp \sqrt{m^2-4\Lambda^2}}{2},\quad y=1 \ee for which
\be
P_2^2(x)- 4 Q_2(x) = (x+m)^2 (x^2 - 4\Lambda^2).
\ee
This shows that both solutions give rise to the same
Seiberg-Witten curve. They are nevertheless inequivalent solutions
for the gauge theory on $\mathbb{R}^3 \times S^1$.

There is one more solution which is not strictly speaking a
solution. It corresponds to the baryonic branch. Consider the
$\epsilon\rightarrow 0$ limit of
\be
p_1=p_2=-m+\epsilon, \quad y=\epsilon/m.
\ee
One can easily check that the $p_i$ and $q_i$ equations of motion
vanish in this limit, and that the Seiberg-Witten curve becomes
\be \label{jaux}
P_2^2(x)- 4 Q_2(x) = (x+m+\Lambda)^2 (x+m-\Lambda)^2.
\ee
However, this solution is at the boundary of the coordinate
system.

The $r=0$ and $r=1$ solutions can be mapped directly to the first
two limits $(2,0)$ and $(1^u,1^u)$. However, the baryonic branch
is more complicated, as this is exactly where the map between the
$(1^u,1^u)$ and $(1^u,1^d)$ degenerates. One can take limits of
the coordinates in $(2,0)$ and $(1^u,1^u)$ so that the
Seiberg-Witten curve takes the form (\ref{jaux}), but in none of
those limits do the equations of motion also vanish. In any case,
the baryonic root appears to be never a bona fide finite solution of the
equations.

\subsubsection{$N_f=1$ and $3$}
For the limits $(1^u,0)$ and $(1^d,0)$ we find the expected $3$ genus zero SW curves. For $(2, 1^d )$ we find 3 different phases ($r=0,1$ and a baryonic root). The $r=1$ branch is triply degenerate, as expected. The baryonic root is located on the edge of the phase space as above. The gauge theory analysis gives in this case $2$ phases in addition to the three found. These are the phases obtained from the Seiberg dual quark potential in equation (\ref{eq:seibergdualquarkpotential}) when the meson matrix is degenerate. The non-degenerate meson matrix leads back to the electric vacua, which we do find here. 

\subsubsection{$N_c=2, N_f=4$}
This is a simpler version of the case studied in \cite{hollowood} and for direct comparison we take the superpotential in this case to be purely quadratic $W(x) = x^2$. The difference is that in that article the flavour masses were all assumed to be unequal and the solution was calculated in the weak coupling limit. We will not present the calculation here, but we find the expected 2 electric branches: two $r=0$ branches and a quadruply degenerate $r=1$ branch.

\subsubsection{Unequal masses, $N_f=2$}
We study this in the limit $(1^u,1^d)$. Since this example is more involved we will elaborate slightly. We start by imposing a superpotential of the form $W(x) = \frac{1}{2} x^2$. The e.o.m. for the momenta are easily solved. We are then left with one equation for the coordinates which, using the transformation $e^{q_1} = b_1, e^{q_2} = b_2$ reads,
\begin{eqnarray}
\frac{{b_2}\,\left( {\Lambda }^2\,{\left( {{b_1}}^2 - {{b_2}}^2 \right) }^2 + {{b_1}}^2\,\left( {b_1}\,{b_2}\,{{m_2}}^2 - \left( {{b_1}}^2 + {{b_2}}^2 \right) \,{m_2}\,{m_3} + {b_1}\,{b_2}\,{{m_3}}^2 \right) \right) }{{{b_1}}^3 - {b_1}\,{{b_2}}^2} = 0.
\label{eq:coordsdifmas}\end{eqnarray}
This yields $5$ solutions, a trivial one ($b_2 =0$) and the four solutions of the quartic equation. The Seiberg-Witten curve for the trivial solution is 
\begin{equation}
y^2_{\textrm{triv}} = \left( x^2 - {\Lambda }^2 + {m_2}\,{m_3} + x\,\left( {m_2} + {m_3} \right)  \right)^2.\end{equation} 
The solution to the quartic equation is not very insightful by itself and we were unable to show the factorization of the Seiberg-Witten curve directly. However, factorization can be shown through a different route\footnote{another route would be to compute the common divisors of the  of the curve expressed in spin chain variables and the remaining equation of motion}. Note that, since we are first and foremost interested in roots of the Seiberg-Witten curve, we can try to find it's double roots when evaluated on the solution for the momenta. We shall use the quadratic part of the numerator of equation (\ref{eq:coordsdifmas}) repeatedly to simplify expressions. Using this, it can be shown that the spectral curve simplifies to
\begin{eqnarray}
y^2 & =&  -4\,{\Lambda }^2\,\left( x + {m_2} \right) \,\left( x + {m_3} \right)  +
  \left(x^2 + \frac{b_1 (m_2 + m_3)}{b_1 + b_2} x + \frac{2 \Lambda^2 b_1 - b_2}{b_1}\right)^2 .
\label{eq:speccurvedifmas}\end{eqnarray}
We first try to find its double roots by studying the cubic equation $\frac{\partial y^2}{ \partial x}=0$. Happily, this equation factorizes into a linear and a quadratic equation. The linear equation has a root at
\begin{equation}
 - \frac{(m_2 + m_3)b_1}{b_1 b_2}
\end{equation}
This root is a candidate for a double root of the spectral curve (\ref{eq:speccurvedifmas}). Plugging this root back into the curve we see that it is indeed a double root of that curve iff equation (\ref{eq:coordsdifmas}) holds. We have therefore found the double root of the spectral curve. Using polynomial division gives the non-singular part, which has the simple form
\begin{equation}
(x )^2 - \frac{4\,{\Lambda }^2\,{b_2}}{{b_1}}.
\end{equation}
Numerical evaluation of the double roots shows that these consists of solutions which connect (at $m_2 = m_3$) to a doubly degenerate $r=1$ branch and the expected two solutions for the $r=0$ branch.

\subsubsection{Cubic potential, $N_f=2$}
We study a superpotential of the form $W(x) = \frac{1}{3}x^3 - a^2 x$,
using the limit $(1^u, 1^d)$. Solving the equations for the coordinates yields three independent solutions. Two of these are related by $\Lambda \rightarrow - \Lambda$. The third one yields two genus one Seiberg-Witten curves of the same form,
\begin{equation}
y^2 = -4\,{\left( m + x \right) }^2\,{\Lambda }^2 +
  {\left( -a^2 + x^2 + 2 \Lambda^2 \right) }^2.
\end{equation}
The extra contribution in the $P$-polynomial exactly cancels the leading term of the $Q$ polynomial to form $W'(x)^2 + f_{n-1}$ just as expected from the field theory analysis in appendix \ref{ap:curvfact}. There are 2 main branches of solutions remaining with two remaining equations in each of them. Examination of these equations shows these have again $3$ sub-branches of solutions. Since the equations are related by $\Lambda \rightarrow - \Lambda$ for the two main branches, we only have to explore one set of these $3$ sub-branches. The first sub-branch gives again the above genus $1$ curve (quadruply degenerate this time). The second sub-branch gives rise to $4$ different genus zero curves and the last one consists of two doubly degenerate $r=1$ curves.

\subsubsection{Cubic potential, $N_f=1$}

We consider the same potential as above and the e.o.m. give in this case 2 main branches of solutions. One main branch consists of two sub-branches. One of them numerically corresponds to 6 genus zero curves, whereas the other sub-branch contains (numerically) a root at approximately $x=-m$. However, we were unable to obtain an analytical expression for these last curves. The remaining main branch corresponds to two genus one curves, as expected.

\subsection{$N_c = 3$}
\subsubsection{$N_f=4$}
We study the limit $(1^u,2,1^d)$, since this case resembles the correct limit for $N_c=2, N_f=2$ and find $2+2+1$ vacua for $r=0,1,2$ respectively. These vacua belong to the electric phase. Furthermore we find the direct (addition map) lift of the $N_c=2$, $N_f=2$ baryonic branch. It appears that we are missing the same (magnetic) vacua here as we did in the $N_c=2$, $N_f=3$ case. The $r=2$ branch yields 6 times the same curve, the two $r=1$ branches four each, and the $r=0$ just has one solution for each curve, as expected from the field theory analysis.

\subsubsection{$N_f=2$}

We find $4$ $r=0$ and $2$ branches of the solution using the (2,0,0) limit. The $r=1$ branch (which has two solutions for each curve) corresponds to the addition map applied to $N_c=2, N_f=0$, since it has the structure $(x+m)^2$ times the curve for $N_c=2, N_f=0$ with the same superpotential. The $(1^u,1^d,0)$ limit shows one extra phase for a solution of the e.o.m. which is on the edge of the phase space ($q_1 \rightarrow \infty$). This fase has Seiberg-Witten curve factorization
\begin{equation}
y^2 = (x+m)^2((x+m)^2+ 2 \Lambda^2)((x+m)^2- 2 \Lambda^2).
\label{wrongcurve}\end{equation}
which is a genus $1$ curve. This is an example of a solution which is not expected from the gauge theory analysis which lies on the boundary of the phase space.

\subsubsection{$N_f=3$}
For this interesting case we checked several different limits. Just as in the $N_c=N_f=2$ case we generically find the expected $3$ $r=0$ and $3$ (with 3 solutions each) $r=1$ electric branches. Only limits which contain a link like in the $N_f=2=N_c$ case, i.e. $(1^u,1^d)$ also produce the baryonic root.

\subsection{$N_c = 4$}
In order to check the multiplication and the addition map we study the four flavoured theory. The multiplication map can be studied by taking the limit $(1^u,1_d,1^u,1^d)$. This gives 4 phases at $r=0$, 4 at $r=1$, and $2$ phases at $r=2$. Actually, the phase at $r=2$ has a $(x+m)^6$ piece in the Seiberg-Witten curve, but if we shift the mass in the superpotential from $\frac{1}{2} x^2 + m x$ to a slightly different mass $\tilde{m}$ we see that only $4$ roots of the Seiberg-Witten curve remain at $x=-m$. In addition we find the multiplication map lift of the baryonic branch. However, in addition to these phases we also find a Seiberg-Witten curve of the form $((x+m)^8 + \Lambda^8)$ and one of the form $(m+y)^4((y+m)^2 - \Lambda^2)((y+m)^2 + \Lambda^2)$ which are genus 1 curves, both very similar to equation (\ref{wrongcurve}). Just as in that case, these solutions are on the edge of the phase space. In fact, the genus $1$ curves seems to be the addition map lift of this 'wrong curve'. We checked also a different limit for this case, which yields the same phases with the exception of the baryonic root. 

\subsection{$N_f \leq N_c$}
We take the limit $(0^{N_c-N_f} (1^u)^{N_f})$, $N_c$ to be larger than to $2$ and the superpotential to be $W(x) = \frac{1}{2} x^2 + m x$. We take the masses to be all unequal. We compute the superpotential,
\begin{equation}
W = \frac{1}{2} \sum_{i=1}^{N_c} p_i^2 +  \sum_{i=1}^{N_f} \Lambda e^{q_{i-1} - q_i} (p_i +m_i) +  \sum_{i=N_f +1}^{N_c} \Lambda^2 e^{q_{i-1} - q_i} + m \sum_{i=1}^{N_c} p_i.
\label{eq:quadUN}\end{equation}
To see this is indeed the correct expression for the quadratic potential note the following: for Lax matrices at a site which contains masses, the constant lower right matrix element only contributes to the trace of the transfer matrix at order $x^{N_c-3}$. Since a quadratic potential is needed that can be calculated from the coefficient of $x^{N_c-2}$ and $x^{N_c-1}$ that element can be safely ignored for calculational purposes. The general form of the quadratic superpotential can then be inferred by comparing to the Toda case. From equation (\ref{eq:quadUN}) we easily obtain and solve the equations of motion for $p$,
\begin{eqnarray}
p_i + m = - \Lambda e^{q_{i-1} - q_i}  \quad i= 1, \ldots, N_f\\
p_i + m = 0 \quad i= N_f + 1, \ldots, N_c
\label{eq:sol1}\end{eqnarray}
Her the variable $q_0$ is defined as $q_N$. The equations of motion for $q$ are a bit more involved,
\begin{equation}
\begin{array}{rclcl}
e^{q_{i-1} - q_i} (p_i +m_i) & = & e^{q_{i} - q_{i+1}} (p_{i+1} +m_{i+1}) & & i= 1, \ldots, N_f-1 \\
e^{q_{N_f-1} - q_{N_f}} (p_{N_f} +m_{N_f})& = &\Lambda e^{q_{N_f} - q_{N_f+1}} & & i= N_f \\
\Lambda e^{q_{i-1} - q_{i}} & = & \Lambda e^{q_{i} - q_{i+1}} & & i= N_f +1 \ldots N_c -1 \\
\Lambda e^{q_{N_c-1} - q_{N_c}}& = &e^{q_{N_c} - q_1} (p_1 +m_1) & & i = N_c,
\end{array}\end{equation}
or using the solution for $p$ and the definition $e^{q_{i-1} - q_{i}} = z_i$,
\begin{equation}
\begin{array}{rclcl}
z_i (m_i - \Lambda z_i - m) & = & z_{i+1} (m_{i+1}- \Lambda z_{i+1} - m) & & i= 1, \ldots, N_f -1 \\
z_{N_f} (m_{N_f}- \Lambda z_{N_f} - m) & = & \Lambda z_{N_f+1} & & i= N_f \\
z_i & = &  z_{i+1} & & i= N_f +1 \ldots N_c -1 \\
\Lambda z_{N_c} & = & z_1 (m_1 - \Lambda z_1 - m) & & i = N_c.
\end{array}\label{eq:sol2}\end{equation}
The variables $z_i$ obey the constraint $\prod z_i =1$. At this point it is important to note that the equations of motion indeed have a solution, up to the constraint. Also, in this case all solutions give finite coordinates. Evaluated on the solutions for the momenta the transfer matrix becomes
\begin{eqnarray}
T(x) = \Lambda^{N} \prod_{i=N_c}^{N_f+1}\left( \begin{array}{cc} \frac{x}{\Lambda}+ \frac{m}{\Lambda} & z_i \\ - 1 &  0 \end{array}\right) \prod_{j=N_f}^{1} \left( \begin{array}{cc} \frac{x}{\Lambda} + z_j + \frac{m}{\Lambda} & z_j (\frac{m_j}{\Lambda} -  z_j - \frac{m}{\Lambda})  \\ - 1 &  z_j \end{array}\right).
\label{eq:transfermatrixsol}\end{eqnarray}
Here we have obtained the variables $z_j$ by inserting `$1$' between Lax matrices $L_i L_{i-1}$ as \begin{equation}
\left( \begin{array}{cc} 1 & 0 \\ 0 &  e^{q_{i-1}} \end{array}\right) \left( \begin{array}{cc} 1 & 0 \\ 0 &  e^{-q_{i-1}} \end{array}\right),
\end{equation}
and realizing we are only interested in the eigenvalues of the transfer matrix. 

\subsubsection{Equal masses}
First take all masses equal to $m$. Then $N_f$ simple quadratic equations have to be solved, which are solved for $N_f$ less than $N_c$ as $z_i = \epsilon_i \sqrt{- z_{N_c}}$, with $\epsilon_i$ equal to either $+1$ or $-1$. The constraint now reads,
\begin{equation}
 i^{N_f} \left(z_{N_c}\right)^{N_c -\frac{N_f}{2}} \prod_{i=1}^{N_f} \epsilon_i  = 1.
\label{eq:constraintNfNc}\end{equation}
For even $N_f$ this constraint can be solved straightforwardly for $z_{N_c}$ for any choice of $\epsilon_i$. We therefore have a total number of $2^{N_f} (N_c - N_f/2)$ solutions. When $N_f$ is odd, we have to square the constraint which gives $z_{N_C} = (-1)^{\frac{1}{2 N_c - N_f}}$. Inserting this back into the constraint then fixes one of the $\epsilon$, arriving at the same number of solutions. On the basis of the field theory analysis we expect $\binom{N_f}{N_f/2} (N_c - N_f/2) + \sum_{r=0}^{N_f/2 -1} \binom{N_f}{r} (2 N_c - N_f)$ solutions for even $N_f$ and  $\sum_{r=0}^{N_f/2 - 1/2} \binom{N_f}{r} (2 N_c - N_f)$ for odd $N_f$, which can easily be shown to be equal to the number of found solutions using $2^k = \sum_{i=0}^{k} \binom{k}{i}$. If $N_f$ equals $N_c$ the above analysis can be repeated by solving everything in terms of $z_i = \epsilon_i |z_{N_c}|$ (note that there are $N-1$ epsilons). Then we see that there are $2 N_c$ solutions by squaring the constraint, and the constraint fixes one of the epsilons for the two possibilities $(z_{N_c})^{N_c} = \pm 1$. For the rest of this subsection we set the mass $m$ for convenience to zero and consider $N_f < N_c$ since the case $N_f = N_c$ is a simple generalization of the argument below. The mass can be restored by shifting $x \rightarrow x+m$ and shifting $p \rightarrow p-m$ in the solution.

The above counting argument gives the total number of solutions for all $r$ branches. Below we show how to obtain the factorization of the curve. Use the solution of the $z_j$ in terms of the $z_N$ and (cyclically) redistribute factors of $\sqrt{z_{N_c}}$ in (\ref{eq:transfermatrixsol}) to get,
\begin{eqnarray}
T(x) = \Lambda^{N} \prod_{i=N_c}^{N_f+1}\left( \begin{array}{cc} \frac{x}{\Lambda} &  \sqrt{z_{N_c}} \\ -  \sqrt{z_{N_c}} &  0 \end{array}\right) \prod_{j=N_f}^{1} \left( \begin{array}{cc} \frac{x}{\Lambda} + i \epsilon_j \sqrt{z_{N_c}} & \sqrt{z_{N_c}} \\ - \sqrt{z_{N_c}} &  i \epsilon_j \sqrt{z_{N_c}} \end{array}\right).
\end{eqnarray}
Define $\tilde{x} \equiv \frac{x}{\Lambda \sqrt{z_{N_c}}}$, so
\begin{eqnarray}
T(\tilde{x}) = (\sqrt{z_{N_c}} \Lambda)^{N}\tr \prod_{i=N_c}^{N_f+1}\left( \begin{array}{cc} \tilde{x}  &  1 \\ -  1 &  0 \end{array}\right) \prod_{j=N_f}^{1} \left( \begin{array}{cc} \tilde{x} + i \epsilon_j & 1 \\ - 1 &  i \epsilon_j \end{array}\right).
\end{eqnarray}
This expression for the transfer matrix consists of building blocks
\begin{equation}
B \equiv \left( \begin{array}{cc} \tilde{x}  &  1 \\ -  1 &  0 \end{array}\right)
\quad A^+ \equiv \left( \begin{array}{cc} \tilde{x} + i  & 1 \\ - 1 &  i  \end{array}\right)
\quad A^- \equiv \left( \begin{array}{cc} \tilde{x} - i  & 1 \\ - 1 &  - i \end{array}\right)
\end{equation}
These matrices share a basis of eigenvectors and commute. In fact,
\begin{equation}
A^+ \cdot A^- = \tilde{x} B
\label{handyrelation}\end{equation}
This immediately leads to the identification of the $r$-branches in this solution: we take a set of $N_f -r$ $A^+$'s and $r$ $A^-$'s, $r \leq N_f$, $z_N$ the appropriate solution of (\ref{eq:constraintNfNc}) and apply (\ref{handyrelation}) to get the transfer matrix
\begin{equation}
T(\tilde{x}) = (\sqrt{z_{N_c}} \Lambda)^{N} (\tilde{x})^r B^{N_c -N_f +r} (A^\pm)^{N_f+r},
\end{equation}
which will yield a spectral curve of the form $y^2 \sim \tilde{x}^{2 r} G(\tilde{x})$. The $\pm$ depends on whether $r\leq [\frac{N_f}{2}]$ or not. From equation (\ref{eq:constraintNfNc}) we see that we have $N_c- \frac{N_f}{2}$ solutions of this type for even $N_f$ and $r\leq \frac{N_f}{2}$. Therefore for $r < \frac{N_f}{2}$ there are $2 \binom{N_f}{r} (N_c - \frac{N_f}{2})$ solutions for each $r$. The factor of two comes from the fact that for fixed $r$ both the solutions with $r$ $A^+$'s as the solutions with $r$ $A^-$ contribute. For $r = \frac{N_f}{2}$ there are therefore only $N_c- \frac{N_f}{2}$ solutions. For odd $N_f$ we have to take into account that there has to be either one $A^+$ or $A^-$ more than the other. Then we have $\binom{N_f}{r} (2 N_c - N_f)$ solutions at each $r$-level, so the counting of states works out level by level.  

Now writing the building block matrices in an eigenvalue basis gives a transfer matrix with diagonal elements:
\begin{eqnarray}
T_{11}(\tilde{x}) = (\sqrt{z_{N_c}} \Lambda)^{N}  \left(\frac{\tilde{x}}{2} + \sqrt{(\frac{\tilde{x}}{2})^2 - 1}\right)^{N_c-N_f} \left(\frac{\tilde{x}}{2} + i + \sqrt{(\frac{\tilde{x}}{2})^2 - 1}\right)^{N_f-r}  \nonumber \\
 \qquad \left(\frac{\tilde{x}}{2} -i+ \sqrt{(\frac{\tilde{x}}{2})^2 - 1}\right)^{r}\\
T_{22}(\tilde{x}) = (\sqrt{z_{N_c}} \Lambda)^{N}  \left(\frac{\tilde{x}}{2} - \sqrt{(\frac{\tilde{x}}{2})^2 - 1}\right)^{N_c-N_f} \left(\frac{\tilde{x}}{2} + i - \sqrt{(\frac{\tilde{x}}{2})^2 - 1}\right)^{N_f-r}  \nonumber \\
 \qquad \left(\frac{\tilde{x}}{2} -i - \sqrt{(\frac{\tilde{x}}{2})^2 - 1}\right)^{r} .
\end{eqnarray}
The square of the difference between the diagonal elements is the discriminant of the characteristic equation of the $2$ by $2$ matrix, so is the Seiberg-Witten curve,
\begin{equation}
y^2 = (T_{11}-T_{22})^2.
\end{equation}
By first reducing by using (\ref{handyrelation}) it is easily seen that the remaining Seiberg-Witten curve has the form polynomial times $\sqrt{(\frac{\tilde{x}}{2})^2 - 1}$ and thus factorizes.

Motivated by \cite{kennawaywarner} we show how to write this in term of linear combinations of Chebyshev polynomials. Chebyshev polynomials of first and second kind obey the identities
\begin{eqnarray}
T_n(x) &=& \frac{1}{2}\left(\left(x +  \sqrt{(x^2-1)}\right)^n + \left(x - \sqrt{(x^2-1)}\right)^n \right) \\
U_{n-1}(x) \sqrt{x^2-1} &=& \frac{1}{2}\left(\left(x +  \sqrt{(x^2-1)}\right)^n - \left(x -  \sqrt{(x^2-1)}\right)^n \right).
\end{eqnarray}
It is easily seen by expanding
\begin{equation}
\left(\frac{\tilde{x}}{2} + a \pm \sqrt{(\frac{\tilde{x}}{2})^2 - 1}\right)^{r} = \sum_{k=0}^{r} \binom{r}{k} a^{r-k} \left(\frac{\tilde{x}}{2}  \pm \sqrt{(\frac{\tilde{x}}{2})^2 - 1}\right)^{k},
\label{eq:chebexp}\end{equation}
that the trace of the transfer matrix is a linear combination of Chebyshev polynomials of the first kind, while the Seiberg-Witten curve is a linear combination of Chebyshev polynomials of the second kind times the expected non-singular piece. 

\subsubsection{Unequal masses}
In this case far less specific information than above can be obtained since to obtain more concrete results in this case the constraint equation $\prod z_i = 1$ has to be solved, which is hard to do in general. However, note that inserting the solution for the $z_i$ in term of $z_{N_c}$ into equation (\ref{eq:transfermatrixsol}) leads to a transfer matrix of the form 
\begin{eqnarray}
T(x) = \Lambda^{N} \prod_{i=N_c}^{N_f+1}\left( \begin{array}{cc} \frac{x}{\Lambda} + \frac{m}{\Lambda}& z_{N_c} \\ - 1 &  0 \end{array}\right) \prod_{j=N_f}^{1} \left( \begin{array}{cc} \frac{x}{\Lambda} + \tilde{z}_j + \frac{m}{\Lambda} & z_{N_c} \\ - 1 &  \tilde{z}_j \end{array}\right),
\end{eqnarray}
where $\tilde{z}_j$ are the roots of 
\begin{equation}
z_j (\frac{m_j}{\Lambda} -  z_j - \frac{m}{\Lambda}) = z_{N_c}.
\label{eq:zisolutions}\end{equation}
Set $\hat{x} \equiv \frac{x +m}{\Lambda \sqrt{z_{N_c}}}$ and $\hat{z_j} = \frac{\tilde{z_j}}{\sqrt{z_{N_c}}}$ to obtain
\begin{eqnarray}
T(x) = \left(\sqrt{z_{N_c}} \Lambda\right)^{N} \prod_{i=N_c}^{N_f+1}\left( \begin{array}{cc} \hat{x} & 1 \\ - 1 &  0 \end{array}\right) \prod_{j=N_f}^{1} \left( \begin{array}{cc} \hat{x} + \hat{z_j}  & 1 \\ - 1 &  \hat{z_j} \end{array}\right),
\end{eqnarray}
The building block matrices in this transfer matrix all commute so can be diagonalized simultaneously, as before. The eigenvalues now differ from previous cases only by the value of $a$ in equation (\ref{eq:chebexp}) since the differences in the Lax matrices between the cases with and without equal masses are proportional to the identity matrix. Hence we can (again) write the spectral curves of these transfer matrices in terms of Chebyshev polynomials and the spectral curve factorizes in the appropriate manner,
\begin{equation}
y^2_{\textrm{diff mass}} = H^2 \sqrt{\left(\frac{\hat{x}}{2}\right)^2 -1}.
\end{equation}
From the point of view of gauge theory it is interesting that we find a concrete realization of the polynomials studied in \cite{kennawaywarner} generating the factorization for both equal and unequal masses for the full range of $N_f \leq N_c$.

\subsection{Massive vacua}
In line with the calculation in \cite{firstpaper} we show how to generalize the above computation to the massive vacua of an arbitrary (single trace) polynomial superpotential for $N_f \leq N_c$. Since the factorization into a genus zero curve shown above is very special the solution of the integrable system equations of motion shown is a natural starting point to look for a solution for all polynomial superpotentials $W = \sum^{n+1}_{k=1} g_k \frac{x^k}{k}$. From the gauge theory point of view for massive vacua the Seiberg-Witten curve must factorize into a genus zero curve which we take to be 
\begin{equation}
P^2 - 4 Q = H^2 \left((\left(\frac{x+c}{\sqrt{z_{N_c}}\Lambda}\right)^2 - 4 \right).
\end{equation}
According to equation (\ref{eq:curvfact}) there must then be polynomials $G$ and $f$ such that
\begin{equation}
\left(\left(\frac{x+c}{\sqrt{z_{N_c}}\Lambda}\right)^2 - 4 \right) G_{n-1}^2(x) = W'_n(x)^2 + f_{n-1}(x)
\label{eq:nonsingpart}\end{equation}
holds. Since this factorization problem is exactly the same as in the pure glue case which was analyzed in \cite{firstpaper} we only state here that for such polynomials to exist, the constant c must be such that the coefficient $c_0$ in 
\begin{equation}
W'(\sqrt{z_{N_c}} \Lambda (\xi + \frac{1}{\xi}) -c) = \sum_{i=-n}^{n} c_i \xi^i
\label{eq:finalequation}\end{equation}
is zero. We will first show below that a slight generalization of that solution indeed solves the equations of motion up to one equation which we then show to be equal to the equation expected from the gauge theory. 

We take our solution of the equations of motion in (\ref{eq:eqofmotiongen}) to be equations (\ref{eq:sol1}) and (\ref{eq:sol2}) together with the constraint $\prod z_i =1$, where in equation (\ref{eq:sol1}) we replace $m$ by the variable $c$. We have already shown that solutions of this type yield factorized Seiberg-Witten curves with non-singular part as in equation (\ref{eq:nonsingpart}), but we still have to show that these solutions actually solve the equations of motion. 

First change variables as $x = \sqrt{z_{N_c}} \Lambda \tilde{x} -c$ in the integral of equations (\ref{eq:eqofmotiongen}) above. The equations of motion for the $q_i$ are satisfied because the appropriate derivative of the trace of the transfer matrix evaluated on the equation of motion gives (after some manipulations)
\begin{equation}
\frac{\delta T_N}{\delta q_i} = \sqrt{z_{N_c}} \Lambda \left(\begin{array}{cc} 0 & -\sqrt{z_{N_c}}\\ - \sqrt{z_{N_c}} & 0 \end{array}\right) T_{N-1},
\label{eq:tracetransfafg}\end{equation}
where $T_{N-1}$ is the transfer matrix obtained by eliminating Lax matrix $L_i$, evaluated on the solution to the equation of motion. In particular, $T_{N-1}$ is diagonalized a known basis of eigenvectors. In this basis the other matrix becomes
\begin{equation}
 \left(\begin{array}{cc} 0 & -\sqrt{z_{N_c}} \Lambda\\ - \sqrt{z_{N_c}}\Lambda & 0 \end{array}\right) \rightarrow  \left(\begin{array}{cc} 0 & \frac{1}{2}(x + \sqrt{x^2 - 4}) \\ \frac{1}{2}(x - \sqrt{x^2 - 4}) & 0 \end{array}\right).
\end{equation} 
Therefore the trace of equation (\ref{eq:tracetransfafg}) vanishes and the e.o.m. for the q's are satisfied.

The e.o.m. for the momenta can be derived from the transfer matrix
\begin{equation}
\frac{\delta T_N}{\delta p_i} =  \left(\begin{array}{cc} -1 & z_i\\ 0 & 0 \end{array} \right) T_{N-1}.
\label{eq:tracetransfafg2}\end{equation}
Changing basis to the basis of eigenvectors of the other Lax matrices in the transfer matrix (and some manipulations) yields
\begin{equation}
\left(\begin{array}{cc} -1 & \hat{z}_i \\ 0 & 0 \end{array} \right) \rightarrow \left(\begin{array}{cc} \frac{\hat{z}_i + \frac{1}{2}(\tilde{x} - \sqrt{\tilde{x}^2-4})}{\sqrt{\tilde{x}^2-4}} & \frac{\hat{z}_i + \frac{1}{2}(\tilde{x} + \sqrt{\tilde{x}^2-4})}{\sqrt{\tilde{x}^2-4}}\\ \frac{-\hat{z}_i + \frac{1}{2}(- \tilde{x} + \sqrt{\tilde{x}^2-4})}{\sqrt{\tilde{x}^2-4}} & \frac{-\hat{z}_i + \frac{1}{2}(- \tilde{x} - \sqrt{\tilde{x}^2-4})}{\sqrt{\tilde{x}^2-4}} \end{array}\right).
\end{equation}
Therefore the trace of equation (\ref{eq:tracetransfafg2}) becomes
\begin{equation}
\tr \frac{\delta T_N}{\delta p_i} = \frac{(2 \hat{z}_i +\tilde{x})(T^{11}_{N-1} - T^{22}_{N-1}) - \sqrt{\tilde{x}^2 -4}(T^{11}_{N-1} + T^{22}_{N-1})}{\sqrt{\tilde{x}^2-4}},
\label{eq:geneqeom}\end{equation}
with
\begin{equation}
T_{N-1} = \left( \begin{array}{cc} T^{11}_{N-1} & 0\\0 &T^{22}_{N-1} \end{array} \right).
\end{equation}
Now we can also express the Seiberg-Witten curve on the solution in terms of the transfer matrix $T_{N-1}$ using the fact that on that solution
\begin{equation}
T_N =  \sqrt{z_{N_c}}\Lambda \left(\begin{array}{cc} \frac{\hat{z}_i + \frac{1}{2}(\tilde{x} - \sqrt{\tilde{x}^2-4})}{\sqrt{\tilde{x}^2-4}} & 0 \\ 0 & \frac{\hat{z}_i + \frac{1}{2}(\tilde{x} + \sqrt{\tilde{x}^2-4})}{\sqrt{\tilde{x}^2-4}} \end{array}\right) \left( \begin{array}{cc} T^{11}_{N-1} & 0\\0 & T^{22}_{N-1}\end{array} \right).
\end{equation}
As explained above, the Seiberg-Witten curve is the difference between the diagonal values, which gives something proportional to equation (\ref{eq:geneqeom}). We therefore arrive at
\begin{equation}
\left. \frac{\frac{\delta P}{\delta p_i}}{\sqrt{P^2 - 4 Q}}\right|_{\textrm{solution} } = \frac{1}{\sqrt{z_{N_c}}\Lambda \sqrt{\tilde{x}^2-4}}.
\end{equation}
Hence the equations of motion for all the momenta collapse into $1$ equation. Combining this computation with the shift in $x$ and changing variables $x \rightarrow \xi + \frac{1}{\xi}$ yields
\begin{equation}
\oint W'(\sqrt{z_{N_c}} \Lambda(\xi + \frac{1}{\xi}) -c) \frac{1}{\xi} d\xi = 0,
\end{equation}
which is equivalent to equation (\ref{eq:finalequation}) expected from the gauge theory point of view. We have therefore proven that this particular solution solves the equations of motion for the integrable system and is such that equation $(\ref{eq:curvfact})$ holds. Note that this holds for all $N_f \leq N_c$ and for arbitrary masses. 

As a technical aside, note that in the Toda case the derivatives of the transfer matrix with respect to the momenta can be summed such that on the solution the following equation holds: 
\begin{equation}
\sum_i \frac{\partial P}{\partial p_i} = -\frac{\partial P}{\partial x}.
\end{equation}
This reproduces, when evaluated in the resolvent, exactly the condition in \cite{firstpaper} that $\tr W'(M)$ should vanish, where $M$ is the periodic Toda chain Lax operator.

It is instructive to realize at this point that the factorized Seiberg-Witten curves obtained in this subsection should be the full field theory answer to the factorization problem for massive vacua of gauge theories with fundamental matter with $N_f \leq N_c$. In particular, they account for all possible vacua in a unified way. However, the above construction is constructive in the sense that we do not prove that these are all possible factorizations. On the other hand, they are all possibilities that are expected from the weak coupling point of view by a simple counting argument. To compare to other approaches to gauge theory we calculate in the final subsection the vevs for the adjoint scalar in these vacua. 

\subsection{Adjoint scalar vevs for massive vacua}
The vacuum expectation values for the adjoint scalar in the massive vacua discussed above are given by 
\begin{equation}
<\!u_k\!> = -\frac{1}{2 \pi i} \oint x^{k-1} \log\left(\frac{(P(x) + \sqrt{P^2 - 4 Q})}{x^{N_c}} \right) dx .
\end{equation}
For the massive vacua
\begin{equation}
P(x) + \sqrt{P^2 - 4 Q} = 2 T_{11} = 2 (\sqrt{z_{N_c}}\Lambda)^{N_c} \prod_{i=1}^{N_c} \left(\frac{\tilde{x}}{2} + z_i + \sqrt{\left(\frac{\tilde{x}}{2}\right)^2 -1} \right) ,
\end{equation}
with $\tilde{x} = \frac{x+c}{\sqrt{z_{N_c}} \Lambda}$. The expectation values can be split into a part which also appears in the pure glue case and extra terms depending on the flavours,
\begin{eqnarray}
<\!u_k\!>  & = & <\!u_k\!>_{\textrm{pure glue}} +<\!u_k\!>_{\textrm{matter}} \\
<\!u_k\!>_{\textrm{pure glue}} & =&  -\frac{1}{2 \pi i} \oint x^{k-1} N_c \log\left(\frac{\sqrt{z_{N_c}}\Lambda \left( \frac{\tilde{x}}{2} + \sqrt{\left(\frac{\tilde{x}}{2}\right)^2 -1}\right)}{{x}} \right)dx  \\
<\!u_k\!>_{\textrm{matter}} & =&  -\frac{1}{2 \pi i} \oint x^{k-1}    \log \left(1 + \frac{z_i}{\frac{\tilde{x}}{2} + \sqrt{\left(\frac{\tilde{x}}{2}\right)^2 -1}}   \right) dx.
\end{eqnarray}
Calculating the glue term first we get, by splitting the logarithm into 
\begin{equation}
\log \left(1 + \frac{c}{x} \right) + \log\left(\frac{1}{2} + \sqrt{\frac{1}{4} - \frac{1}{\tilde{x}^2}}\right),
\label{eq:someintegral}\end{equation}
two finite integrals which can be performed. The first vev is easily calculated to be 
\begin{equation}
<\!u_1\!>_{\textrm{pure glue}} =  -c \,N_c .
\end{equation}
The needed contour integral of both logarithms in equation (\ref{eq:someintegral}) can be evaluated, with the help of the expansion\footnote{Take a derivative on the left hand side w.r.t. $y \equiv \frac{1}{x}$ and observe that this is $- \frac{C(y)}{\sqrt{1-4y}}$ with $C(y)$ the generating function for the Catalan numbers}
\begin{equation}
\log\left(\frac{1}{2} + \sqrt{\frac{1}{4} - \frac{1}{\tilde{x}^2}}\right) = - \sum_{i=1}^{\infty} \frac{1}{2 i} \binom{2 i}{i} \frac{1}{\tilde{x}^{2i}},
\end{equation}
to give
\begin{equation}
<\!u_k\!>_{\textrm{pure glue}} =  \frac{N_c}{k} \sum_{l=0}^{\left[\frac{k}{2}\right]} \binom{k}{2 l} \binom{2l}{l} (z_{N_c} \Lambda^2)^{l} (-c)^{k- 2 l} .
\label{eq:gluecontrib}\end{equation}
The matter contribution can be calculated by using the change of variables $\tilde{x} \rightarrow \xi + \frac{1}{\xi}$. As an intermediate result 
\begin{equation}
-\frac{1}{2 \pi i}\oint_{\xi = \infty} (\xi + \frac{1}{\xi})^j (1-\frac{1}{\xi^2})\log\left(1 + \frac{z_i}{\xi}\right) = \sum_{m=0}^{\left[\frac{j}{2}\right]} \frac{1}{j-m+1} \binom{j}{m} (-z_i)^{j-2m+1}
\end{equation}
is obtained by straightforward computation. Plugging this result into the needed integral we arrive at
\begin{equation}
<\!u_k\!>_{\textrm{matter}} = (\sqrt{z_{N_c}} \Lambda)^k \sum_{i=1}^{N_f} \sum_{j=0}^{k-1}  \sum_{m=0}^{\left[\frac{j}{2}\right]}  \binom{k-1}{j} \binom{j}{m} \frac{(-z_i)^{j-2m+1}}{j-m+1}  \left(\frac{-c}{\sqrt{z_{N_c}} \Lambda}\right)^{k-1-j} .
\label{eq:matcontrib}\end{equation}
The full answer for the adjoint scalar vev in the massive vacua is now the sum of equations (\ref{eq:gluecontrib}) and (\ref{eq:matcontrib}). Comparing above expressions to the ones obtained by a matrix model calculation in \cite{janikdemasure} we immediately identify 
\begin{eqnarray}
c \rightarrow - T & \quad & z_{N_c} \Lambda^2 \rightarrow R .
\end{eqnarray}
Their constraint is exactly the constraint $\prod z_i = 1$. This can be seen by multiplying the two different solutions to equation (\ref{eq:zisolutions}) for each site with a mass. For the solutions of that equation a consistent sign has to be chosen to connect to the matrix model result. With these identifications the relation to equations (67), (68), (69) and (70) in \cite{janikdemasure} is obvious. The relation with the remaining equality (71) is more obscure, but for $k=2$ there is perfect agreement. It would certainly be interesting to check the matrix model calculation further against the above field theory result. 

\subsection{Toward a general proof of factorization}
In this final subsection we indicate how a proof of factorization of the Seiberg-Witten curve can be obtained up to a technical detail. In previous work \cite{secondpaper} we relied on a theorem in \cite{mumfordmoerbeke} which relates straight line flows on the Jacobian of the spectral curve to the equations of motion of the periodic Toda chain. This connection is lacking in the case of the classical spin chain, although the fact that it is integrable and periodic strongly suggests that the reasoning should hold in exactly the way. In this paper however, we are mainly interested in the critical points of the potential as can be obtained by using equation (\ref{eq:eqofmotiongen}), reproduced here for convenience,
\begin{equation}
\frac{\delta W}{\delta o_i} = \frac{-1}{2 \pi i} \oint W'(x) \frac{\frac{\delta P}{\delta o_i}}{\sqrt{P^2 - 4 Q}}. \nonumber \end{equation}
These equations can be written as
\begin{equation}
\frac{\delta W}{\delta o_i} = \frac{1}{2 \pi i} \sum_j \frac{\delta s_j}{\delta o_i} \oint W'(x) \frac{x^{N_c-j}}{\sqrt{P^2 - 4 Q}}. \end{equation}
Assuming that the transformation Jacobian is not trivial, the equations of motion amount to the vanishing of certain one forms on the Riemann-surface, a condition actually identical to the conditions derived in \cite{secondpaper}. Therefore, modulo the technical assumption, this amounts to a proof of factorization as in equation (\ref{eq:curvfact}) of the spectral curve of the integrable system in exactly the same way as derived from the gauge theory. 

\section{Integrating in the mesons}
\label{sec:integrmesons}
A natural question in this framework is whether or not one can integrate in the mesons. A natural reason to suspect that this might be the case is the fact that the solutions we found for $N_f < N_c$ are calculationally very similar to vacua of the potential (\ref{potentialnflessnc}). The gauge theory interpretation of the solutions of both superpotentials is also similar. In addition, pure field theory arguments indicate that it might be possible to integrate in the mesons and even the baryons \cite{intriligator94}. Finally, in \cite{yves} it was shown that the minima of the ADS potential and the minimum of the matrix model potential agree for a quadratic potential. In this section we show how to directly integrate in the meson fields for a quadratic potential in the case $N_f \leq N_c$ and how this reproduces both four and three dimensional results.

The quadratic \xxx superpotential (obtained from equation (\ref{eq:quadUN}), reproduced here for convenience) is for the limit $1^u$ on the first $N_f$ sites
\begin{equation}
W = \mu \left( \sum_{i=1}^{N_c} p_i^2 + 2 \sum_{i=1}^{N_f} \Lambda e^{q_{i+1} - q_i} (p_i +m_i) + 2 \sum_{i=N_f +1}^{N_c} \Lambda^2 e^{q_{i+1} - q_i} \right). \nonumber 
\end{equation}
Here $q_{N+1}$ is identified with $q_1$. A convenient way to rewrite this is in
terms of new variables $y_i = \Lambda e^{q_{i+1}-q_i} (p_i + m_i)$. These satisfy one constraint
which we impose using a Lagrange multiplier $S$, and we get:
\begin{equation}
W =  \mu  \sum_{i=1}^{N_c} (p_i^2 + 2 y_i) + S \log \left(\frac{\Lambda^{2 N_c - N_f} \prod^{N_f}_{j=1} (p_j+m_j)}{\prod y_j} \right).
\label{eq:84}\end{equation}
In this equation $\Lambda$ only appears in the logarithm and it couples
 to $S$. Therefore, $S$ is to be identified with the familiar glueball
field of pure ${\cal N}=1$ theories. We now introduce new Lagrange multipliers $M_i$ for the masses and new fields $\tilde{m}_i$ and we rewrite (\ref{eq:84}) as
\begin{equation}
W =  \mu \sum_{i=1}^{N_c} (p_i^2 + 2 y_i) + S \log\left(\frac{\Lambda^{2 N_c - N_f} \prod_{j=1}^{N_f}(p_j+\tilde{m}_j)}{\prod y_j}\right) + \sum_{i=1}^{N_f} (m_i-\tilde{m_i}) M_i
\label{eq:mesintin}\end{equation}
If we integrate out $M$ we regain (\ref{eq:84}). On the other hand, we
can also integrate out $p_i$, $\tilde{m}_i$ and all $y$ except $y_{N_c}$ and we find
\begin{equation}
W =  \sum_{i=1}^{N_f}\left(- \frac{M_i^2}{2 \mu} + m_i M_i \right)+ 2 \mu y_{N_c} + (N_c-N_f-1)S+ S \log \left(\frac{\Lambda^{2 N_c - N_f} 2^{N_c} \mu^{N_c}}{y_{N_c} S^{N_c-N_f-1} \prod M_j}\right).
\label{eq:branchpoint}\end{equation}
Integrating out $S$ gives
\begin{equation}
W =  \sum_{i=1}^{N_f}\left(- \frac{M_i^2}{2 \mu} + m_i M_i \right)+ 2 \mu y_{N_c} + (N_c-N_f-1) \left(\frac{\Lambda^{2 N_c - N_f} 2^{N_c} \mu^{N_c}}{y_{N_n}\prod M_j}\right)^{\frac{1}{N_c-N_f-1}}.
\end{equation}
From now on we identify in the above expression $\prod M_j$ with the determinant of the meson matrix. In order to compare to the three-dimensional $\mathcal{N} = 2$ theories studied in \cite{aharonyseiberg}, we want to take the limit $\mu \rightarrow \infty$, keeping $\Lambda^{2 N_c - N_f} \mu^{N_c}$ fixed by scale matching. Furthermore we set the quark masses to zero and arrive at the superpotential for the four dimensional gauge theory we study, compactified on a circle \cite{aharonyseiberg},

\begin{equation}
W = Y + (N_c-N_f-1)\left(Y \det M\right)^{\frac{1}{N_c-N_f-1}},
\end{equation}
with a convenient normalization. If $N_f = N_c-1$, then integrating out S in the final step yields the right quantum constrained moduli space.

Another direction from equation (\ref{eq:branchpoint}) is integrating out $y_{N_c}$ and $S$, which yields equation (\ref{potentialnflessnc}) upon putting the masses to zero and trivially redefining $\mu$ as $2 \mu$. In this case we can also treat $N_f = N_c$ and arrive at the correct quantum constrained moduli space. Baryons can be integrated in by using the procedure described in one of the examples in \cite{intriligator94}, but this does not seem to yield illuminating results.

When $N_c < N_f < 2 N_c$ in principle one can also apply equation (\ref{eq:mesintin}) by redefining the $y$ variables appropriately. However, it is difficult to obtain a concrete general expression for the superpotential in this case. For the low order cases we checked (up to $N_c =3$) the answer is exactly the potential expected for the electric branches of the theory. We hope to return to this issue elsewhere. 

\section{Discussion}
We have shown in this paper that the \xxx spin chain provides good holomorphic coordinates for part of the moduli space of $U(N)$ gauge theory with $N_f$ fundamentals compactified on $\mathbb{R}^3 \times S^1$. Some coordinate patches touch the baryonic root at the edge of phase space, but there are sometimes other solutions to the equations of motion on this edge as well. These other solutions do not have a gauge theory interpretation. Being on the edge neither class is a bona fide vacuum of our superpotential, but it is certainly suggestive and it would be interesting to understand this further. Since the coordinates provide a coordinate patch for part of the moduli space it is natural to wonder whether there is a more general integrable system which provides coordinate patches for the other parts of moduli space, or even coordinates that are valid everywhere. In particular it would be interesting to find a system which includes the baryonic root holomorphically. It might be worthwhile to see if there is a more general formulation of the integrable system starting from the formulation in $SL(2, \mathbb{C})$ variables. 

We have also shown that it is natural to integrate in the mesons into the superpotential by Legendre transforming with respect to the masses. The baryonic operators cannot be naturally integrated in and it remains an open question whether or not there is an extension of the integrable system in which there is a parameter multiplying the baryon. This parameter would then correspond to the variable $b$ in \cite{intriligator94} and could be used to integrate in the baryon. More generally it remains a question whether the meson and dyon vevs have an interpretation in the integrable system. An avenue of approach to this problem starts with noticing that the meson vevs are simply the value of the matrix model resolvent at $x=-m_i$. This gives at least on shell a method of backtracking to the integrable system what these vevs are. 

Yet another open problem remains the microscopic interpretation of the phase space variables $p_i$ and $q_i$ in the three dimensional gauge theory. A way to examine this issue could be to study perturbative instanton contributions in the gauge theory to find the coordinate transformation of our results to the right microscopic variables, just as was used in \cite{firstpaper}.

As indicated before, we expect that the factorization of the Seiberg-Witten curve on the minima of our superpotentials which is expected on field theory grounds can be proven along the lines of the second proof in \cite{secondpaper}. We have indicated above how such a proof works in principle, but it would be interesting to close the small technical gap in that proof. Interesting generalizations of results in this paper include but are not limited to studying other gauge groups, for which the integrable system is, to our knowledge, unknown and/or to higher dimensional gauge theories.

\vspace{.5in}

{\bf Acknowledgments: }
We would like to thank Asad Naqvi, Leo Kampmeijer, Jeroen Wijnhout and Robert Duivenvoorden for useful discussions. This work is partially supported by the Stichting FOM.

\appendix
\section{Heisenberg variables}\label{sec:heisvar}

In this appendix we will rewrite (\ref{poissonalg}) in term of the Heisenberg variables for which $\{q_i,p_i\}=1$. The aim will be to obtain something which resembles equation (\ref{todachain}). We will drop the indices on the $q$'s, $p$'s and $S$'s to declutter notation. First rewrite the algebra as,
\begin{eqnarray}\label{poissonalgacc}
\{S_0, S_{\pm}\} = \pm S_{\pm} \qquad \{S_+, S_-\} = 2 S_0
\end{eqnarray}
We will begin with the Ansatz $S_0 \equiv -p -\lambda_i$ and read (\ref{poissonalgacc}) as a system of differential (derivational) equations. The first can be easily solved by taking,
\begin{eqnarray}
S_{\pm} = f_{\pm}(p) e^{\pm q}
\end{eqnarray}
where the functions $f_{\pm}$ contain a constant of integration. Putting this into the second equation we obtain,
\begin{equation}
f_+(p) e^{-q}\{e^q, f_-(p)\}  + e^q f_-(p) \{f_+(p), e^{-q}\} = 2 (- p - \lambda_i)
\end{equation}
Take the functions $f_{\pm}$ to be polynomials of degree $2$, $f_{\pm} = a_{\pm} + b_{\pm} p + c_{\pm} p^2$. This yields three different solutions,
\begin{eqnarray}\label{heisvar}
& &\left\{  \begin{array}{ccl} f_- & = & a_ -  \frac{2 \lambda_i}{a_+} p - \frac{1}{a_+} p^2 \\ f_+ &=& a_+ \end{array} \right. \\
& &\left\{  \begin{array}{ccl} f_- &=& a_- + b_- p  \\ f_+ &=& (- 2 \frac{\lambda_i}{b_{-}} + \frac{a_-}{b_{-}^2}) - \frac{1}{b_-} p  \end{array}\right. \\
& &\left\{ \begin{array}{ccl} f_- &=& a_- \\ f_+ &=& a_+ -  \frac{2 \lambda_i}{a_-} p - \frac{1}{a_-} p^2 \end{array} \right.
\end{eqnarray}
The constants can be connected to $\lambda_i$ and $K_i$ through $S_0^2 + S_+ S_- = K_i^2$, which yields,
\begin{eqnarray}
& &\left\{\begin{array}{ccl} f_- &=&-\frac{\lambda_i^2- K_i^2}{a_+} -  \frac{2 \lambda_i}{a_+} p - \frac{1}{a_+} p^2 \\ f_+ &=& a_+ \end{array} \right. \\
& &\left\{\begin{array}{ccl} f_- &=& b_-(\pm K_i + \lambda_i) + b_- p  \\ f_+ &=& -( \frac{\mp K_i + \lambda_i}{ b_-}  )- \frac{1}{b_-} p  \end{array}\right. \\
& &\left\{\begin{array}{ccl} f_- &=& a_- \\ f_+ &=&  \frac{K_i^2 - \lambda_i^2}{a_-} -  \frac{2 \lambda_i}{a_-} p - \frac{1}{a_-} p^2 \end{array} \right.
\end{eqnarray}
where in the second line 2 solutions are possible (which differ only in which mass is in which off-diagonal entry).

\section{Seiberg-Witten curve factorization in the strong coupling approach}
\label{ap:curvfact}

In this appendix it is shown how to calculate factorization loci of the Seiberg-Witten curve when turning on a classical superpotential in the gauge theory approach, particularly when the superpotential is of high degree ($W \sim \tr \Phi^{n+1}$, $n \geq 2N_c - N_f$).  In this appendix single trace superpotentials are studied for $N_f < 2 N_c$. 

The theory at the root of the $r$-th Higgs
branch \cite{argyresseiberg} is a $U(r) \times U(1)^{N_c-r}$ gauge
theory, with $N_f$ massless hypermultiplets charged under the
$U(1)$ subgroups of $U(r)$. On special points along the root $l-r$
extra massless hypermultiplet degrees of freedom appear which are
charged under the $U(1)$ subgroups. At these points the
Seiberg-Witten curve factorizes as 
\begin{equation}
P(x)^2 - 4 Q(x) = (x-m)^{2r} \prod_{i=1}^{l-r}(x-p_i)^2  \prod_{j=1}^{2 N- 2l-2r} (x-q_j) \equiv H_l(x)^2 F_{2N-2l}(x).
\end{equation}
Here $P(x) = \sum_{i=0}^{N_c} s_i x^{N_c -i}$ and $Q \equiv \Lambda^{2
N_c - N_f} \left(\sum_{i=0}^{N_f} \tilde{m}_i x^{N_f -i} \right)$. 

We will denote the extra massless dyons (and their scalar
components) by $d_i$ and $\tilde{d}_i$. Note that since $r$
eigenvalues of $\Phi$ are fixed we are (classically) left with a
$U(N_c-r)$ gauge group. For this theory we can write the superpotential
\begin{equation}
W = \sqrt{2} \sum_{i=1}^{l-r} \tilde{d}_i a_i d_i + \sum_{k=1}^{n+1}
g_k u_k. \label{eq:strcouplpot}\end{equation} Since the theory is
classically (for a generic point on the Higgs branch part of the
vacuum) broken to a $U(N_c -r)$ gauge theory with $N_f - 2 r$
flavours, the $u_k$ correspond to multi-trace operators for $k >
N_c-r$. Therefore, for $n < N_c-r$ the analysis below works\footnote{For $n=N_c -r$ we do not have to add dyons in the case the Seiberg-Witten curve does not degenerate further. In this case the superpotential can be minimized by writing it first in terms of the $N_c-r$ eigenvalues.  Then they can be taken to be equal to the $N-r$ different roots of $W'$ such that the non-singular part of the curve is $W_{N_c -r}'(x)^2 + f_{N_f - 2r}$. For $n \leq 2 N_c - N_f $ this has the expected form of $W'(x) + f_{n-1}$. A more refined argument based on the gauge theory resolvent is also possible.}.

There is an ambiguity in the definition of the operators $u_k$,
since for higher values of $k$ there can be several quantum operators with the same
classical limit. We will denote the `classical' $u_k$ by
$u^{(c)}_k$ in the following. In other words, $u_k^{(c)}= \frac{1}{k} \sum_i
a_i^k$ for $P(x)=\prod_i(x-a_i)$. However, for the $u_k$ appearing
in the superpotential (\ref{eq:strcouplpot}) we will use the
quantum operators whose vacuum expectation values are obtained
from the gauge theory resolvent. The quantum vevs
$u^{(q)}_k$ can be expressed as
\begin{equation}
u^{(q)}_k = \frac{1}{2 \pi i} \oint_{v=0}dv v^{-k-1}
\left(\ln \left(P(1/v)\right) +
\ln \left(1 + \sqrt{1 - 4 \frac{Q(1/v)}{P(1/v)^2}}\right) \right).
\end{equation}
The first logarithm gives for $k \leq N_c$ exactly the Newton-Girard
relations between power sums and the $s_i$, which can be written as
symmetric polynomials of the eigenvalues. The second term produces
the quantum corrections. Clearly, in the limit $\Lambda\rightarrow
0$, $Q\rightarrow 0$ and $u^{(q)}_k\rightarrow u_k^{(c)}$ as expected. We define $u^{(q)}_k = u_k^{(c)} + \Delta u_k$. 

Before proceeding to the general case, we first calculate the first few corrections. Since only residues at $v=0$ are needed, the quantum contributions $\Delta u_k$
from the second logarithm can be expanded as 
\begin{eqnarray}
\Delta u_k =  \frac{1}{2 \pi i} \oint_{0}dv v^{-k-1}\left( v^{2
N_c - N_f} \frac{v^{N_f} Q(1/v)}{(v^{N_c}P(1/v))^2} + \frac{3}{2}
v^{4 N_c -2 N_f} \left(\frac{v^{N_f}
Q(1/v)}{(v^{N_c}P(1/v))^2}\right)^2 + \right. \nonumber \\
  \left. \phantom{\left(\frac{v^{N_f}
Q(1/v)}{(v^{N_c}P(1/v))^2}\right)^2} +  \mathcal{O}(v^{6 N_c - 3N_f}) \right). \qquad  \label{eq:uquantumdif}\end{eqnarray} 
%there is a slightly improper use of a phantom in the previous equation: the term in the phantom physically should not be there, but makes the formulae work out nicely....
The fraction in the integrand can now be Taylor-expanded around $v=0$ since it
does not contain poles any more. The quantum $u^{(q)}_k$ and the
classical $u^{(c)}_k$ differ only when $k \geq 2 N_c - N_f$ and the first two corrections are
\begin{eqnarray}
k=2 N_c - N_f    & & \Delta u_k = \Lambda^{2 N_c-N_f} \label{eq:difftrphisq} \\
k=2 N_c - N_f +1 & & \Delta u_k = \Lambda^{2 N_c-N_f}\left(\tilde{m}_1  +2 s_1 \right) - \frac{3}{2} \Lambda^{4 N_c - 2 N_f} \delta_{2 N_c - N_f, 1}. \label{eq:deltauquant}
%k=2 N_c - N_f +2 & & u^{(q)}_k = u^{(c)}_k + \Lambda^{2 N_c-N_f}\left(\tilde{m}_2 - 2 \tilde{m}_1 s_1 + 3 s_1^2 - 2 s_2\right) + 3 \Lambda^{4 N_c - 2 N_f} \left(\tilde{m}_1 -2 s_1 + \frac{10}{9}\Lambda^{2 N_c - N_f}\right)\delta_{(2 N_c - N_f), 1} + \frac{3}{2} \Lambda^{4 N_c - 2 N_f} \delta_{(2 N_c - N_f), 2}
\end{eqnarray}

The point of this calculation is that it matters which
superpotential is used in (\ref{eq:strcouplpot}) in a calculation
along the lines of \cite{deboeroz} to calculate Seiberg-Witten
curve factorization loci. We will do this below for a (single trace) superpotential derived from the gauge theory resolvent. Equation (\ref{eq:strcouplpot}) may be varied with respect to $u^{(q)}_k$ to obtain these. However, in that calculation we would need $\frac{\delta s_j}{\delta
u^{(q)}_k}$. This is hard to calculate since the $s_j$ are highly
non-trivial functions of the $u^{(q)}_k$. It is easier to vary equation (\ref{eq:strcouplpot}) with respect to the classical moduli, since part of the calculation is then standard. This yields 
\begin{equation}
- \sum_{l=0}^{n+1} g_l \frac{\delta(u^{(c)}_l +  u^{(q)}_l)}{\delta
u^{(c)}_k} = \sqrt{2} \sum_{i=1}^{N_c-r} \tilde{d}_i d_i
\frac{\delta a_i}{\delta u^{(c)}_k}.
\end{equation}
Repeating the steps in \cite{deboeroz} gives,
\begin{equation}
\tilde{W}_n'(x) = B_{q} \frac{P(x)}{H_l(x)} + \mathcal{O}(\frac{1}{x}),
\label{eq:deboeroztype}\end{equation}
with
\begin{equation}
\tilde{W}'(x) \equiv  \sum_{r=1}^{n+1}g_r x^{r-1} +  \sum_{l=0}^{n+1} \sum_{r=1}^{n+1} g_l \frac{\delta u^{(q)}_l}{\delta u^{(c)}_r} x^{r-1}.
\label{eq:deftildew}\end{equation} $B_{q}$ is a
polynomial\footnote{in \cite{deboeroz} this polynomial was denoted
by $H(v)$} of degree less than or equal to $l-1$ and to get the neat first term in $\tilde{W}'(x)$ we have used the assumption that the degree of the potential is less than or equal to $N_c$. By comparing
leading powers in $x$ in equation (\ref{eq:deboeroztype}), it follows that $N_c-n \leq l \leq N_c-1$. Squaring equation (\ref{eq:deboeroztype}) now gives an equation akin to equation
$(3.44)$ in \cite{balastonaqvi},
\begin{equation}
\tilde{W'}^2_n(x) = B_{q}^2 F_{2n - 2q} + f_{n-1}(x) +
\frac{4 Q(x) B_{l-1}^2}{H^2_{2 l}(x)} +
\mathcal{O} (\frac{1}{x}). \label{eq:factorizationlocus}\end{equation}
This equation indicates that $B_{q}^2 F_{2n - 2q} = W_n'^{(c)}(x)^2 + g_{2n - 2 N_c + N_f}$ with $g$ a polynomial of the indicated order. From other considerations we expect to find in the end $W'^{(c)}(x)^2 + \tilde{f}_{n-1}$ and below we show that this is actually the case if the extra contributions on both left and right hand side of this equation are taken into account. 

The left hand side of the equation can be rewritten in terms of the classical superpotential and its quantum contributions using equation (\ref{eq:deftildew}). Using the obvious definition $\tilde{W}' \equiv W'^{(c)}_n +W'^{(q)}$ the left hand side of the equation reads
\begin{equation}
(W'^{(c)})^2  + 2 W'^{(c)} W'^{(q)}(x) + (W'^{(q)}(x))^2.
\end{equation}
Note that the extra terms on the right hand side appear at the same order as those on the left hand side. The extra terms on the left can be calculated by starting with
\begin{equation}
\frac{\delta u^{(q)}_l}{\delta u^{(c)}_r} = \frac{-1}{2 \pi i} \oint_{x'=\infty} x'^{l-1} \frac{1}{\sqrt{1-\frac{4 Q}{P^2}} + 1-\frac{4 Q}{P^2}} \left(\frac{4 Q}{P^2} \right)\frac{\tilde{P}}{P} dx',
\end{equation}
where $\tilde{P}$ denotes $\frac{\delta P_r}{\delta u^{(c)}_r}$. The last term in the integrand can be expressed as,
\begin{equation}
\frac{\tilde{P}}{P} = - \frac{\sum_{j=0}^{N_c} s_{j-r} x^{N_c-j}}{\sum_{k=0}^{N_c} s_k x^{N-k}},
\end{equation}
which gives after some massaging,
\begin{equation}
\frac{\tilde{P}}{P} = - x^{-r} (1 - \frac{\sum_{j=N_c-r+1}^{N_c} s_{j} x^{-j}}{\sum_{k=0}^{N_c} s_k x^{-k}}).
\end{equation}
The second term above is suppressed in the integrand by a factor of $x^{-N_c-1}$ and will therefore not contribute to the residue integral as long as the order of the superpotential $n$ does not exceed $N_c$. Plugging this result back into $W'^{(q)}(x)$ yields
\begin{equation}
W'^{(q)}(x) =   \sum_{l=0}^{n+1} \sum_{r=1}^{n+1} g_l x^{r-1} \frac{1}{2 \pi i} \oint_{x'=\infty} x'^{l-r-1} \frac{1}{\sqrt{1-\frac{4 Q}{P^2}} + 1-\frac{4 Q}{P^2}} \left(\frac{4 Q}{P^2} \right) dx'.
\end{equation}
For functions of the form $f(x) = \sum_{i=0}^{\infty} a_i x^{p-i}$ 
\begin{equation}
\left.f(x)\right|_+ =   \sum_{i=0}^{p} \frac{x^{p-i}}{2 \pi i} \oint_{x'=\infty} x'^{i-p-1}f(x') dx'
\end{equation}
holds where $\left.f(x)\right|_+ = \sum_{i=0}^{p} a_i x^{p-i}$ and the 'residua at infinity integral' is taken to be the same as in equation (\ref{eq:gaugetheoryresolvent}). Defining $r-1 = n-j$,
\begin{equation}
W'^{(q)}(x) =   \sum_{l=0}^{n+1} g_l \sum_{j=0}^{n} x^{n-j} \frac{1}{2 \pi i} \oint_{x'=\infty} x'^{j-n-1} x'^{l-1} \frac{\frac{4 Q}{P^2}}{\sqrt{1-\frac{4 Q}{P^2}} + 1-\frac{4 Q}{P^2}} dx'
\end{equation}
gives
\begin{equation}
W'^{(q)}(x) =  \left.\left( W'^{(c)}(x) \frac{1}{\sqrt{1-\frac{4 Q}{P^2}} + 1-\frac{4 Q}{P^2}} \left(\frac{4 Q}{P^2} \right) \right)\right|_+.
\end{equation}
Therefore we arrive at an expression for the left hand side of equation (\ref{eq:factorizationlocus})
\begin{equation}
 W'^{(c)}(x)^2 + 2 W'^{(c)}_n W'^{(q)}(x) + (W'^{(q)}(x))^2 =   W'^{(c)}(x)^2 + \left.\left(W'^{(c)}(x)^2 \left(\frac{\frac{4 Q}{P^2}}{1- \left(\frac{4 Q}{P^2} \right)} \right) \right)\right|_+.
\end{equation}
The field theory moduli coordinates on the left hand side of equation (\ref{eq:factorizationlocus}) are the double roots, while on the right hand side these are the $s_i$. These are related by the factorization of the Seiberg-Witten curve and using this the extra term on the right hand side of equation (\ref{eq:factorizationlocus}) is
\begin{equation}
\frac{4 Q(x) B_q^2}{H^2_{2 l}(x)} =  \frac{\frac{4 Q}{P^2}}{1 - \frac{4 Q}{P^2}} (W'^{(c)}(x)^2 + g_{m}).
\end{equation}

In the above equation $m$ denotes the order of the extra polynomial in the factorization formula. To start of it is of order $2n - 2 N_c + N_f$, which is smaller than $2n$. This polynomial order was derived from equation (\ref{eq:factorizationlocus}). However, we see that the leading term which does not depend on the polynomial cancels on both sides, so the polynomial is of one order lower. Then the argument can be repeated until $g$ is of order $n-1$. This completes the argument that even for high order superpotentials the factorization of the Seiberg-Witten curve is captured in equation (\ref{eq:curvfact}) if one consistently uses the gauge theory resolvent to calculate the quantum operators $u_k$. 

\section{An observation}
\label{sec:anobservation}
A standard technique in matrix model calculations for theories with fundamental matter is integrating out the matter content: the resulting matrix model can then be tackled by the usual techniques. The modified potential takes a form like \begin{equation} \sim \log(\det(\Phi + m))\end{equation} for every integrated flavour. Considering that the translation for the pure glue case proceeded by directly identifying the matrix integral variable with the Lax matrix of the periodic Toda chain, we might try to postulate that one has to take the modified matrix model potential and insert the Toda chain Lax matrix into it to obtain the effective action on $\mathbb{R}^3 \times S^1$. This would be in line with the observation in \cite{gomezreino} that matrix model calculated superpotentials for systems with matter in the fundamental can be constructed from data of systems without. Note that in this way we can introduce at most $N_f \leq N_c$ flavours. Also, this kind of potential leads to equations of motion of high order rapidly. We calculated (some of) the minima of a potential $W = m x + \frac{1}{2} x^2 + h \log(\det(x + m))$ for $N_c$ equal to $2$ and $3$ to check this idea. Note that we leave a term $h$ of dimension $2$ in front of the potential undetermined for reasons which will become clear below. In the matrix model this is $N_f$.

\subsubsection*{$N_c=2$}
We plug the periodic Toda chain Lax matrix into the potential. In a full calculation we would have to use the field theory resolvent to calculate the potential correctly, but here we will follow a more direct route: we simply take the $z$-independent part of the determinant before taking the logarithm. Solving the equations of motion we get $6$ curves. Four of these have a double root. In addition, there is a doubly degenerate curve of the form
\begin{equation}
y^2 = ((x+m)^2 + 2 \Lambda^2 + h \Lambda^2)((x+m)^2 - 2 \Lambda^2 + h \Lambda^2)
\end{equation}
Now if this curve is to have double points, $h$ has to be $\pm 2$, where it has a double root at $-m$. The rest curve has the form of $(x+m)^2 \pm 4 \Lambda^2$. Note that the calculation with the \xxx chain gives a minus sign in this curve. The remaining $4$ curves can now be evaluated for the same two values of $h$, and these give 3 different curves since 2 solutions collapse to the same curve solution:
\begin{eqnarray}
y^2 &=& ((x+m)^2 \mp 4 \Lambda^2)(x+m)^2 \\
y^2 &=& (x+m)(m+x+ 4 \sqrt{\mp 1}  \Lambda )(m+x+ 2 \sqrt{\mp 1} \Lambda)^2\\
y^2 &=& (x+m)(m+x- 4 \sqrt{\mp 1}  \Lambda )(m+x- 2 \sqrt{\mp 1} \Lambda)^2
\end{eqnarray}
Note that we do not get the correct form of the curve for the two solution which probably should correspond to the $r=0$ branch. Also, we get one more curve at $r=1$ than is expected from the field theory analysis. To further check this type of construction we turned to $N_c=3$.

\subsubsection*{$N_c =3$}
We first have to fix the coefficient $h$. For this purpose we use the solution with equal momenta $p_1=p_2=p_3$. This gives twelve solutions to the equations of motion, all of which have 2 double roots in the Seiberg-Witten curve (for unfixed $h$ even). This branch has a curve with double roots at $-m$ if and only if $h = 0$. We therefore try to find a $h$ such that it reproduces at least one of the double roots of the $r=1$ branch of the 3 flavoured theory. We find that we can fix $h$ such that 2 of the curves have the correct double roots (but not the single roots), but then we can numerically simply verify that the other curves do not have double roots at a recognizable location. This indicates that this approach to finding the factorization locus of Seiberg-Witten curves with matter in the fundamental in this particular setup does not work. It might be that the solutions of the equations of motion we have not considered do have the correct double roots. However, then one would still have to explain that there is a huge excess of solutions, for which there is no (intrinsic) reason to discard them. We have therefore checked that the constructed potential for three colors cannot (fully) describe a system with 2 flavours, as one might suspect.

\subsubsection*{Discussion}
Although an intriguing observation at the level of two colors, the constructed potential in this particular form obviously does not reproduce four dimensional field theory answers for theories with more than $2$ colors and is therefore not the correct integrable systems-inspired exact superpotential for systems compactified on a circle. However, based this observation we conjecture that there is a (generic for all $N$) non-polynomial superpotential for the periodic Toda chain which reproduces Seiberg-Witten curve factorizations for supersymmetric gauge theories with fundamental matter ($N_f = N_c$) which for 2 colors resembles the superpotential displayed above. Using the field theory resolvent to calculate the non-polynomial potential might be a good starting point to improve the above calculation. In any case, a further study of equilibria of non-polynomial potentials for the Toda chain would certainly be very interesting.

\end{document}